\documentclass[journal]{IEEEtran}
\usepackage{amsmath,amsfonts}
\usepackage{array}
\usepackage[caption=false,font=normalsize,labelfont=sf,textfont=sf]{subfig}
\usepackage{textcomp}
\usepackage{balance}
\usepackage{stfloats}
\usepackage{url}
\usepackage{verbatim}
\usepackage{graphicx}
\usepackage{cite}
\usepackage[linesnumbered,ruled,vlined]{algorithm2e}
\usepackage{xcolor}
\usepackage{booktabs}
\usepackage{multirow}
\usepackage{hyperref}
\hypersetup{
    colorlinks=true,  
    linkcolor=blue,  
    citecolor=blue, 
    urlcolor=black 
}

\newcommand{\rev}[1]{#1}
\newcommand{\revb}[1]{#1}

\begin{document}

\title{Object-Attribute-Relation Model Driven Adaptive Hierarchical Transmission for Multimodal Semantic Communication}

\author{$^{\star}$$^{\S}$$^{\dagger}$Chenxing Li, $^{\star}$$^{\S}$$^{\dagger}$Yiping Duan, $^{\star}$$^{\S}$Han Jiao, $^{\star}$$^{\S}$$^{\dagger}$$^{\ddagger}$Xiaoming Tao, $^{\ast}$Weiyao Lin and $^{\star}$Mingquan Lu \\
$^{\star}$ Department of Electronic Engineering, Tsinghua University, Beijing, China\\
$^{\S}$ State Key Laboratory of Space Network and Communications, Beijing, China \\
$^{\dagger}$ Beijing National Research Center for Information Science and Technology (BNRist), Beijing, China \\
$^{\ddagger}$School of Computer Science and Technology, Xinjiang University, Urumqi, China \\
$^{\ast}$ Department of Electronic Engineering, Shanghai Jiao Tong University, Shanghai, China

\thanks{Copyright \copyright~2026 IEEE. Personal use of this material is permitted. However, permission to use this material for any other purposes must be obtained from the IEEE by sending an email to pubs-permissions@ieee.org.}
\thanks{Xiaoming Tao(taoxm@mail.tsinghua.edu.cn) is the corresponding author. This work was supported in part by the National Natural Science Foundation of China under Grant NSFC 62595731, 62595733, 62595735, 62531012,62322109, U22B2001; in part by the Xplorer Prize in Information and Electronics technologies, and the Nantong Research Institute for Advanced Communication Technologies, and the Program of Jiangsu Province under Grant NTACT-2024-Z-001.}}
\markboth{IEEE Transactions on Circuits and Systems for Video Technology}%
{Li \MakeLowercase{\textit{et al.}}: O-A-R Model Driven Adaptive Hierarchical Transmission for Multimodal Semantic Communication}


\maketitle

\begin{abstract}
Traditional video coding (VVC, HEVC) prioritizes human visual perception, transmitting substantial texture redundancy that severely hinders machine decision-making under constrained bandwidths. \revb{In dynamic channels, this redundancy squanders scarce bandwidth and incurs prohibitive latency, and the paradigm offers no mechanism to prioritize decision-critical semantics as capacity fluctuates.} To address this, we propose a robust multimodal semantic communication framework based on an adaptive Object-Attribute-Relation (O-A-R) hierarchy. Bypassing pixel-level reconstruction entirely, our framework directly fuses visual, textual, and audio streams to construct a decision-oriented \rev{O-A-R graph}. A bandwidth-adaptive strategy dynamically allocates resources by semantic priority, while a cross-modal mechanism leverages text and audio priors to compensate for severe visual degradation. Experimental results demonstrate that under extreme low bandwidths (1-3 kbps), our method achieves over a 90\% bandwidth saving (an approximately 10-fold reduction) compared to state-of-the-art digital schemes, maintaining superior \rev{O-A-R graph} accuracy. In deep fading channels (SNR $\le$ 4 dB), \revb{the proposed hierarchical design ensures} graceful degradation by strictly preserving foundational object anchors even when traditional codecs suffer 100\% decoding failure. Coupled with an 89\% reduction in end-to-end latency, \revb{our framework meets the real-time perception requirements of embodied agents whose decisions consume semantics rather than pixels}.
\end{abstract}


\section{Introduction}
\begin{figure*}[!t]
\centering
\includegraphics[width=\textwidth]{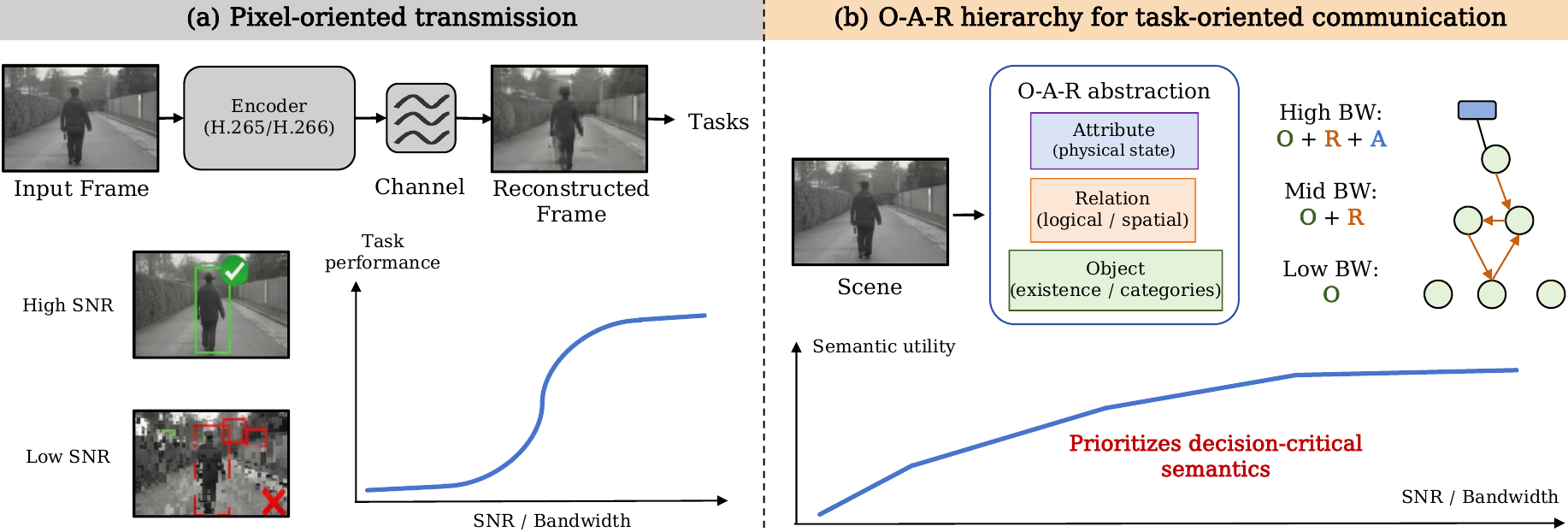}
\caption{\rev{\textbf{Motivation of task-oriented O-A-R semantic communication.} \revb{(a) Pixel-oriented video coding optimizes visual fidelity, spending bandwidth on texture redundancy irrelevant to machine tasks; under low SNR or tight bandwidth budgets its task utility degrades abruptly. (b) O-A-R abstraction transmits task-relevant semantic layers in priority order, preserving object anchors under low bandwidth and degrading gracefully.}}}
\label{fig_1}
\end{figure*}

\IEEEPARstart{W}{ith} the rapid evolution of sixth-generation (6G) mobile networks and Internet of Things technologies, global multimedia data traffic is experiencing explosive growth \cite{ericsson2016mobility, samsung20206g}. Meanwhile, the core paradigm of communication systems is undergoing a fundamental shift from traditional human-to-human interactions to machine-to-machine and human-machine collaboration \cite{gunduz2022beyond}. In emerging embodied intelligence scenarios such as industrial inspection and disaster rescue, the receiver is no longer a human observer demanding high visual fidelity, but an intelligent agent requiring real-time decision-making based on environmental semantics \cite{vuong2023open}. However, existing mainstream video coding standards (H.265/HEVC and H.266/VVC) primarily optimize human visual perception quality, such as minimizing mean squared error or improving structural similarity \cite{bross2021overview}. These methods inevitably transmit massive texture redundancy that contributes little to machine comprehension. Under bandwidth-limited or dynamically fluctuating channel conditions, this redundancy becomes a performance bottleneck restricting real-time inference capabilities (Fig.~\ref{fig_1}(a)) \cite{duan2020video}. Consequently, semantics-centric semantic communication, especially Video Coding for Machines technology, has emerged as a key solution to break the bandwidth barrier and empower efficient intelligent perception \cite{qin2021semantic, du2025object}.

Under such bandwidth-limited, low signal-to-noise ratio (SNR) conditions, traditional video compression introduces artifacts and blurring \cite{bourtsoulatze2019deep} that corrupt image semantics and cause downstream perception algorithms to fail \cite{mao2024survey, o2022impact}. Overcoming this bottleneck requires a compact structured semantic representation (Fig.~\ref{fig_1}(b)). By abstracting data into ``Object-Attribute-Relation'' (O-A-R) triplets, it optimally bridges underlying sensory data and high-level decision logic \cite{krishna2017visual}. Crucially, agent decision-making is inherently hierarchical: \rev{fundamental tasks require only key ``objects'' (Object layer); relational reasoning and planning then demand their logical associations (Relation layer); and finer operations require physical states (Attribute layer). This yields a semantic priority of Object $>$ Relation $>$ Attribute, where object anchors are indispensable nodes, relations encode the decision-critical topology, and attributes are node-level refinements} \cite{gu2024conceptgraphs}. However, existing semantic communication works mostly optimize pixel- or feature-level reconstruction \cite{yang2024swinjscc}, lacking mechanisms to dynamically adapt to these hierarchical semantic requirements under extremely constrained bandwidths.

Furthermore, embodied perception is not restricted to a single visual channel: the agent processes a continuous multimodal stream of visual frames, ambient audio, and textual instructions \cite{baltruvsaitis2018multimodal}. Since real-time decisions hinge on the semantic integrity of the current multimodal snapshot, we optimize the joint transmission of single-frame snapshots as the cornerstone for handling complex video streams \cite{wang2016temporal}. These modalities are strongly complementary---acoustic signatures cue an occluded or poorly lit target, while textual instructions guide attention when textures are blurred \cite{zhao2018sound}. Transmitting raw modalities by linear superposition, however, wastes spectrum through information overlap; an ideal architecture instead performs semantic-level fusion, eliminating mutual information for a ``$1+1+1<3$'' overhead while exploiting cross-modal validation for a ``$1+1+1>3$'' robustness gain.

\rev{Building on the above analysis, this paper proposes a machine-task-oriented multimodal O-A-R hierarchical semantic communication framework that directly decodes a structured O-A-R graph at the receiver~\cite{li2024object}, schedules its transmission by semantic priority under dynamic channels, and fuses visual, textual, and audio streams for cross-modal robustness.}

\revb{\textbf{Working hypothesis and scope.} This work rests on an explicit hypothesis: the receiver is a machine agent whose decision-making consumes a structured semantic description of the scene, rather than a human viewer requiring pixel-accurate reconstruction. Receiver-side tasks of exactly this kind are common in embodied intelligence: an industrial inspection robot acts on the identities, states, and spatial relations of equipment (e.g., ``the valve above the pipe is leaking'') rather than on photographic detail; a rescue robot plans over obstacles, victims, and passable routes; and scene-level visual question answering can be served directly from a symbolic scene description. The adequacy of O-A-R-structured semantics for such reasoning is well documented in the scene-graph literature~\cite{krishna2017visual, gu2024conceptgraphs}, and our zero-shot VQA evaluation in Sec.~IV instantiates such a receiver explicitly---questions are answered solely from the received O-A-R graph, without any image reconstruction. This hypothesis equally delimits the boundary of the framework: it targets semantic-level perception and reasoning, whereas tasks requiring pixel evidence (e.g., photometric inspection or metric-accurate localization) lie outside its intended use.}

The main contributions are summarized as follows:

1. We propose an end-to-end multimodal semantic communication architecture for embodied intelligence. By fusing modalities to directly generate hierarchical O-A-R graphs without image reconstruction, this paradigm eliminates pixel-level redundancy and ensures real-time machine decision-making under constrained bandwidths.

2. We design a bandwidth-adaptive O-A-R hierarchical transmission strategy. By mapping channel states to semantic priorities, the system automatically discards high-level information to protect core object data during channel degradation, achieving \rev{controllable, priority-ordered \emph{graceful degradation} through unequal error protection: as the channel degrades, the foundational object anchors are preserved first}.

3. We introduce a cross-modal semantic compensation mechanism \rev{that leverages textual and audio priors to restore corrupted visual semantics under low SNR or visual occlusion. As quantified in our modality ablation (Table~\ref{tab:ablation_multimodal}), the Text+Audio pathway alone sustains an object-level semantic graph when the visual stream is absent, and cross-modal fusion measurably improves recall in the visually degraded regime}.

\section{Related Work}
\subsection{Deep Joint Source-Channel Coding}
Traditional wireless multimedia transmission is based on Shannon's separation theorem, performing source compression followed by channel coding. This theory holds under assumptions of infinite block lengths and allowable delays; however, it exhibits significant sub-optimality in short block length and low-latency scenarios, and is highly susceptible to the ``cliff effect,'' where performance degrades sharply with slight channel deteriorations \cite{gunduz2024joint}. This issue is particularly prominent in embodied intelligence and real-time visual communication tasks.

In recent years, deep joint source-channel coding (Deep JSCC) has effectively alleviated these problems by directly mapping source to channel symbols via end-to-end neural networks. The DeepJSCC framework proposed by Bourtsoulatze et al. achieved end-to-end wireless image transmission for the first time \cite{bourtsoulatze2019deep}. Subsequent studies on feedback mechanisms, MIMO adaptive designs, and video JSCC have further enhanced robustness and transmission efficiency over complex channels \cite{kurka2020deepjscc, tung2022deepwive, wu2024deep, li2025star, xiao2026robust}. Compared to traditional separate schemes, these methods achieve smoother performance degradation and more stable visual quality under low SNR conditions.

Meanwhile, Deep JSCC research is gradually shifting from ``human-vision-oriented reconstruction'' to ``task-oriented semantic transmission.'' Recent works have begun to directly transmit deep features supporting downstream tasks rather than high-fidelity pixel reconstructions \cite{fu2024generative, eldeeb2024multi, samarathunga2024autoencoder, han2025few}. However, most existing methods rely on unstructured deep feature representations, offering limited semantic interpretability and difficult to flexibly layer for transmission based on bandwidth variations. In contrast, this paper adopts the O-A-R structured semantic representation, which inherently possesses interpretability and hierarchical scalability. \rev{Prior O-A-R communication work~\cite{du2025object} employs the graph as an auxiliary prior on top of a conventional digital stack (entropy coding, LDPC, and QAM). \revb{The suitability of the O-A-R representation for \emph{dynamic-bandwidth} transmission}, by contrast, stems from a structural property of the graph itself: it is decomposable into nested, monotonically informative tiers (objects $\subseteq$ objects$+$relations $\subseteq$ the full graph), so a single representation can be truncated to any prefix of tiers without re-encoding. This nesting makes the achievable semantic fidelity a graceful, non-increasing function of the number of delivered tiers, providing exactly the unequal-error-protection structure required for variable-length transmission as the channel state $\beta$ varies---a property that unstructured feature- or pixel-based codes, optimized for a single fixed reconstruction, lack.}

\rev{\textbf{Relation to classical and task-oriented JSCC.} The proposed framework performs joint source--channel mapping, and---like recent task-oriented JSCC~\cite{fu2024generative, eldeeb2024multi}---it optimizes downstream task fidelity rather than signal mean-squared error\revb{, and, like the JSCC family at large, it inherits the analog, cliff-free degradation behavior. Neither this inherited robustness nor the bypassing of pixel reconstruction is therefore by itself our point of departure}. The essential distinction is structural and twofold. (i)~\emph{Decomposable semantic-priority tiering with unequal error protection.} Both classical JSCC and existing task-oriented JSCC learn a single end-to-end code optimized for one operating point; in contrast, our symbol stream is explicitly priority-ordered (Object $>$ Relation $>$ Attribute), and the O-A-R graph decomposes into nested, monotonically informative tiers that can be truncated to any prefix and adaptively matched to the channel state---yielding native variable-length transmission and unequal error protection that a monolithic code cannot provide. (ii)~\emph{Cross-modal source aggregation.} The source is a fused multimodal aggregate (vision, text, audio) compressed jointly before channel mapping, exploiting cross-modal redundancy that single-source JSCC cannot. These two properties also distinguish our scheme from prior O-A-R communication~\cite{du2025object}, which carries the graph over a conventional digital stack rather than as a decomposable joint source--channel code. \revb{Throughout this comparison, ``semantic'' denotes the structural fidelity of the recovered O-A-R graph---the correctness of its objects, attributes, and relations, measured by graph-level recall and edit distance (Sec.~IV-A)---as distinct from pixel fidelity or the accuracy of a single fixed task head.} Table~\ref{tab:jscc_comparison} summarizes the comparison.}

\begin{table}[!t]
\centering
\color{black}
\caption{\textbf{Relation to classical and task-oriented JSCC.} ``Objective'' is the training target\revb{: pixel-level MSE, a fixed downstream-task loss, or---ours---the structural fidelity of the O-A-R graph (``semantic'')}.}
\label{tab:jscc_comparison}
\small
\begin{tabular}{lccc}
\toprule
Property & Classical & Task-orient. & \multirow{2}{*}{Ours} \\
 & JSCC & JSCC & \\
\midrule
Objective              & MSE & task     & semantic \\
Reconstructs source    & Yes & No       & No  \\
Semantic-priority UEP  & No  & No       & Yes \\
Adaptive variable rate & No  & No       & Yes \\
Multimodal source      & No  & Rare     & Yes \\
\bottomrule
\end{tabular}
\end{table}

\subsection{Video Coding for Machines and Task-Oriented Semantic Representations}
As machine vision increasingly becomes the primary consumer of multimedia traffic, traditional video coding targeting human perception is fundamentally sub-optimal for machine analytics under constrained bandwidths. Driven by this paradigm shift, Video Coding for Machines (VCM) standardization has gained significant momentum \cite{gao2021video}. Recent advancements have evolved from ROI coding and deep feature compression to end-to-end task-oriented representation transmission, demonstrating potential to significantly reduce communication overhead while maintaining downstream task accuracy \cite{yang2024video, kim2023end, lorkiewicz2025video}.

To further push compression boundaries, structured semantic representations---such as object-relation graphs---have emerged as highly compact alternatives to unstructured pixel or feature grids. By abstracting visual scenes into topological nodes and directed edges, these representations inherently filter out task-irrelevant background noise and texture redundancy. Although structured semantics have been widely explored in high-level computer vision, their potential for adaptive resource allocation and unequal error protection in fading wireless channels remains severely underexplored.

In the communication domain, recent efforts have integrated structured O-A-R semantics into video transmission. Du et al. proposed an O-A-R-based semantic communication framework to assist generative video reconstruction, achieving notable bitrate savings \cite{du2025object}. However, such paradigms still fundamentally rely on reconstructing visual frames, incurring prohibitive computational latency and lacking channel-adaptive transmission mechanisms. \rev{Within this venue, the most closely related efforts either target machine-oriented compression---the standardized pixel codec underlying our strongest digital baseline~\cite{bross2021overview} and learnable multi-scale feature compression for VCM~\cite{kim2023end}, both of which still emit an entropy- and channel-coded bitstream and thus remain cliff-prone---or pursue orthogonal semantic-domain goals: embedding hidden information within a semantic channel~\cite{10711846}, and building optical-flow sketch representations for compact video coding~\cite{10379647}. None addresses bandwidth-adaptive, priority-ordered hierarchical transmission of a multimodal O-A-R graph.} In contrast, our framework abandons pixel reconstruction entirely. By mapping multimodal data directly into a prioritizable O-A-R stream, it enables adaptive hierarchical transmission orchestrated by channel state conditions, making it exceptionally resilient for real-time embodied intelligence in hostile wireless environments.

\revb{Security and integrity of semantic transmission form a further emerging line within this venue that complements our reliability focus. Recent TCSVT studies embed covert information directly in the semantic channel via image semantic steganography~\cite{10711846}, and protect transmitted image semantics against tampering and channel distortion through dual-stage robust semantic watermarking~\cite{huo2026drsw}. These efforts safeguard \emph{who} may read, embed, or verify the delivered semantics, whereas this paper addresses \emph{how much} task-critical semantics survives band-limited fading channels; the two directions are complementary, and the priority-ordered O-A-R stream could itself serve as a carrier for such integrity mechanisms in future systems.}

\subsection{Multimodal Semantic Communication and Cross-Modal Priors}
Recent advances in semantic communication have begun exploring multi-source information integration to improve robustness. While traditional paradigms process visual, textual, and audio streams independently, projecting heterogeneous data into a unified semantic space can effectively exploit cross-modal correlations \cite{radford2021learning, girdhar2023imagebind}. Multimodal signal aggregation has shown immense potential in enhancing perception reliability, particularly in noisy, occluded, or complex physical environments \cite{li2024unified, li2024stnet, wei2025audio, xue2025ad, huang2025enhancing, yin2025cross, wojcik2024case, zhangdynamic, zhang2025bayesian}.

The inherent complementarity among sensory modalities provides a critical foundation for robust signal recovery. Highly compressed textual cues and ambient audio signatures can serve as powerful semantic priors to guide visual understanding when the primary visual channel suffers severe degradation \cite{su2024scanformer, liu2025m2ist, yin2025cross}. These findings indicate that multi-source alignment can significantly enhance the theoretical limits of semantic fidelity and system robustness.

From a communication perspective, most semantic transmission systems remain vision-centric \revb{\cite{bourtsoulatze2019deep, yang2024swinjscc, tung2022deepwive}. Non-visual streams, where they are supported at all, are typically carried over separate conventional links rather than fused with the visual stream before channel mapping.} This paradigm fails to exploit non-visual modalities as cross-channel side information to compensate for deep visual fading. To address this, we introduce a unified cross-modal compensation mechanism. By deeply aggregating multi-source signals, our framework leverages lightweight audio and textual streams to recover corrupted visual representations, achieving robust transmission under constrained bandwidths. Section III details the proposed architecture and resource allocation methodology.

\section{System Architecture and Methodology}
\subsection{Overview}

\begin{figure*}[!t]
\centering
\includegraphics[width=\textwidth]{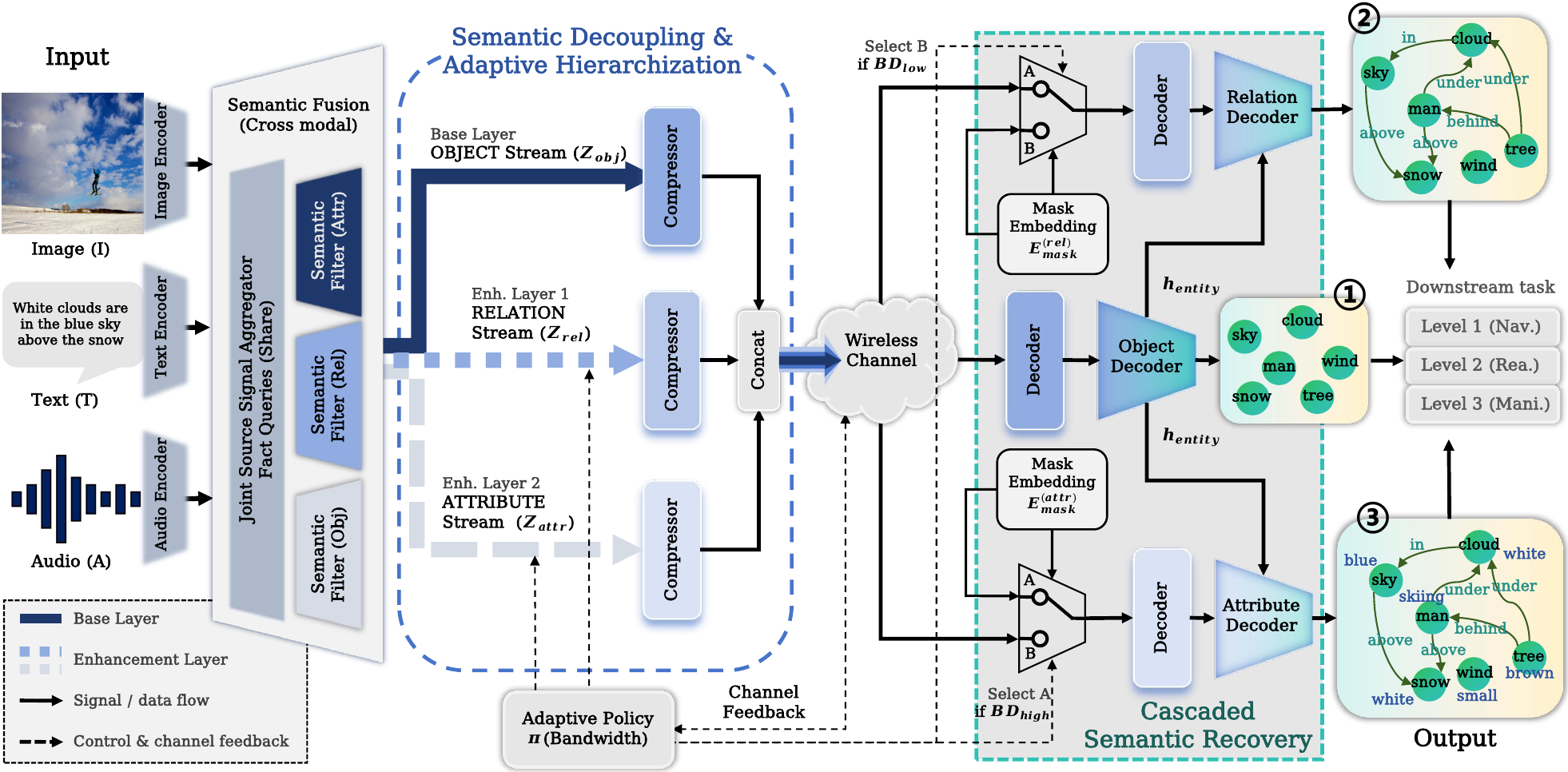}
\caption{\textbf{Block diagram of the proposed multimodal hierarchical semantic communication system.} Multimodal inputs are fused and explicitly decoupled into prioritized object, relation, and attribute streams. An adaptive gating mechanism selectively compresses and transmits these streams based on instantaneous CSI. Finally, a cascaded receiver directly reconstructs the structured \rev{O-A-R graph} for downstream embodied tasks, entirely bypassing pixel-level reconstruction.}
\label{fig_2}
\end{figure*}

This paper proposes a multimodal semantic communication framework based on a hierarchical Object-Attribute-Relation (O-A-R) representation (Fig.~\ref{fig_2}). Designed for embodied agents in dynamic, bandwidth-limited scenarios, it abandons conventional pixel-level reconstruction in favor of a task-oriented semantic transmission paradigm. The framework directly extracts, multiplexes, and transmits decision-critical O-A-R triplets from multi-source sensory inputs. The overall system comprises three core subsystems: the semantic transmitter, the adaptive channel transmission mechanism, and the semantic receiver. \rev{Throughout this paper we use a single term, the \emph{O-A-R graph}; it is equivalent to what prior work variously refers to as a scene graph or a semantic topological graph.}

\revb{\textbf{Scope and assumptions.} As hypothesized in the Introduction, the receiver requires only the O-A-R graph rather than a pixel-level reconstruction of the source; the framework accordingly targets \emph{semantic-level} perception and reasoning---scene understanding, relational reasoning, and visual question answering---rather than pixel-accurate geometry. The graph is category-level: nodes carry object and attribute classes and edges carry relational predicates, the majority of which are spatial/directional (their recovery is quantified in Sec.~IV); absolute metric localization lies outside the present scope and is discussed as future work.}

\textbf{Semantic Transmitter.} This subsystem compresses high-dimensional heterogeneous perceptual data into compact semantic symbols through two sequential stages. 
\begin{itemize}
    \item \textbf{Multimodal Signal Extraction and Aggregation:} Modality-specific front-ends process the raw image, text, and audio streams, while a cross-modal joint source encoder aggregates these complementary signals into a unified latent representation. 
    \item \textbf{Semantic Decoupling and Compression:} The aggregated representation is explicitly decoupled into independent object, attribute, and relation semantic streams via orthogonal routing filters. A semantic compressor then maps these feature spaces into low-dimensional channel symbols to strictly meet stringent wireless bandwidth constraints.
\end{itemize}

\textbf{Adaptive Hierarchical Transmission Mechanism.} To combat time-varying wireless channel fading and capacity uncertainties, we design a bandwidth-adaptive strategy that dynamically allocates transmission resources based on semantic priorities. \revb{As depicted in Fig.~\ref{fig_2}, the instantaneous channel state is estimated at the receiver and returned to the transmitter over a low-rate feedback link (``Channel Feedback'' in Fig.~\ref{fig_2}); the policy controller $\pi(\beta)$ then maps the fed-back bandwidth budget $\beta$ to a transmission mask, and the negotiated mode is shared by both ends so that the variable-length symbol stream is multiplexed and demultiplexed consistently.} The strategy operates in three modes:
\begin{itemize}
    \item Level 1 (Object-Only, Low BW): Retains only fundamental object nodes to guarantee basic agent survival and obstacle avoidance under severe channel degradation.
    \item Level 2 (Object + Relation, Mid BW): Allocates additional bandwidth to relation streams, supporting complex spatial and logical interactions.
    \item Level 3 (Full O-A-R, High BW): Transmits complete attribute details for fine-grained manipulation when channel conditions are highly favorable.
\end{itemize}
After statistical power normalization, the encoded sequence $x$ is transmitted over an Additive White Gaussian Noise (AWGN) channel, yielding the received signal $y = x + n$.

\textbf{Semantic Receiver.} The receiver robustly recovers the underlying \rev{O-A-R graph} from the noisy observations through a cascaded decoding architecture.
\begin{itemize}
    \item \textbf{Signal Unpacking and Completion:} It unpacks the received signal blocks based on the negotiated bandwidth mode, employing a zero-padding strategy for untransmitted layers to maintain dimensional consistency and suppress noise hallucination. 
    \item \textbf{Semantic Decoding and Recovery:} The decoder maps the noisy channel symbols back to the semantic space. Exploiting structural dependencies, cascaded receivers first decode the object anchors, which subsequently serve as contextual side information for recovering the conditionally dependent attribute and relation streams. 
    \item \textbf{O-A-R Graph Generation:} Finally, a structured O-A-R graph is seamlessly generated for downstream planning algorithms, completely bypassing the massive computational latency of traditional video frame reconstruction.
\end{itemize}

\subsection{Semantic Transmitter: Multi-Source Perception and Feature Decoupling}

The semantic transmitter encodes multi-source sensory data into a compact, hierarchical symbol space via three stages: signal extraction, joint-source aggregation, and stream-decoupled compression.

\subsubsection{Multi-Source Signal Extraction}

Three parallel front-ends process visual, textual, and acoustic signals, projecting them onto a unified bandwidth-compatible dimension $D = 256$.

\textbf{Visual Signal Front-end:} A vision-specific extractor \cite{liu2021swin} captures multi-scale spatial semantics from an image $I \in \mathbb{R}^{3 \times H \times W}$. The output undergoes a $1 \times 1$ convolutional projection and 2D positional encoding ($\mathrm{PE}_{2D}$) addition to preserve geometric coordinates, yielding sequence $F_v\in \mathbb{R}^{L_v \times D}$:
\begin{equation}
F_v = f_v(I) = \mathrm{LN}(\mathrm{Conv}_{1 \times 1}(\mathcal{E}_{vision}(I)) + \mathrm{PE}_{2D}) 
\end{equation}
where $L_v = \frac{H}{32} \times \frac{W}{32}$, and $\mathrm{LN}(\cdot)$ denotes LayerNorm for signal stabilization.

\textbf{Textual \& Acoustic Front-ends:} A lightweight sequence encoder \cite{devlin2019bert} tokenizes the instruction $T$ and linearly projects hidden states to define the text feature $F_t \in \mathbb{R}^{L_t \times D}$. Concurrently, the ambient audio $A$ is converted to a Mel-spectrogram, processed by a convolutional acoustic front-end \cite{he2016deep}, and combined with 1D positional encoding to obtain $F_a \in \mathbb{R}^{L_a \times D}$.

\subsubsection{Joint-Source Aggregation and Semantic Decoupling}

To facilitate differential transmission over fading channels, we fuse and explicitly decouple the multi-source signals into independent semantic streams.

\textbf{Joint-Source Aggregation:} Modality signals are concatenated with learnable embeddings ($E_{type}$) to construct a unified memory matrix $M$:
\begin{equation}
M = \mathrm{Concat}([F_v + E_v, F_t + E_t, F_a + E_a])
\end{equation}

\textbf{Semantic Decoupling via Routing Filters:} Transmitting the aggregated $M$ uniformly prohibits Unequal Error Protection (UEP). To enable prioritized transmission, we introduce orthogonal routing filters. Using learnable semantic queries $Q \in \mathbb{R}^{N \times D}$ ($N=30$), $M$ is projected into three independent sub-spaces:
\begin{equation}
z_\tau = \Phi_\tau(\mathcal{H}_{agg}(Q, M; \Theta_{shared}))
\end{equation}
where $\mathcal{H}_{agg}(\cdot)$ is the shared joint-source attention mechanism, and $\Phi_\tau(\cdot)$ is the linear routing filter specific to stream $\tau \in \{obj, attr, rel\}$. This explicitly divides the multimodal signal into focused streams ($z_{obj}$ for basic localization, $z_{attr}$ for state details, $z_{rel}$ for logical interactions).

\subsubsection{Rate-Constrained Semantic Encoding}

To strictly satisfy wireless bandwidth constraints, a weight-shared Semantic Encoder compresses the decoupled vectors $z_\tau$ into a compact channel symbol space. Through non-linear dimensionality reduction, it achieves an 8-fold symbol rate reduction from $D=256$ to $D_c=32$:
\begin{equation}
\mathbf{x}_\tau = \mathrm{Linear}_{128 \to 32}(\sigma(\mathrm{Linear}_{256 \to 128}(\mathrm{LN}(z_\tau))))
\end{equation}

The resulting symbol sequences $\mathbf{x}_{obj}, \mathbf{x}_{attr}, \mathbf{x}_{rel} \in \mathbb{R}^{N \times D_c}$ serve as the foundational transmission units. The encoder's LayerNorm stabilizes the symbol distribution, facilitating accurate signal power normalization before physical antenna transmission.

\subsection{Adaptive Hierarchical Transmission Mechanism}

\begin{figure}[!t]
\centering
\includegraphics[width=3.4in]{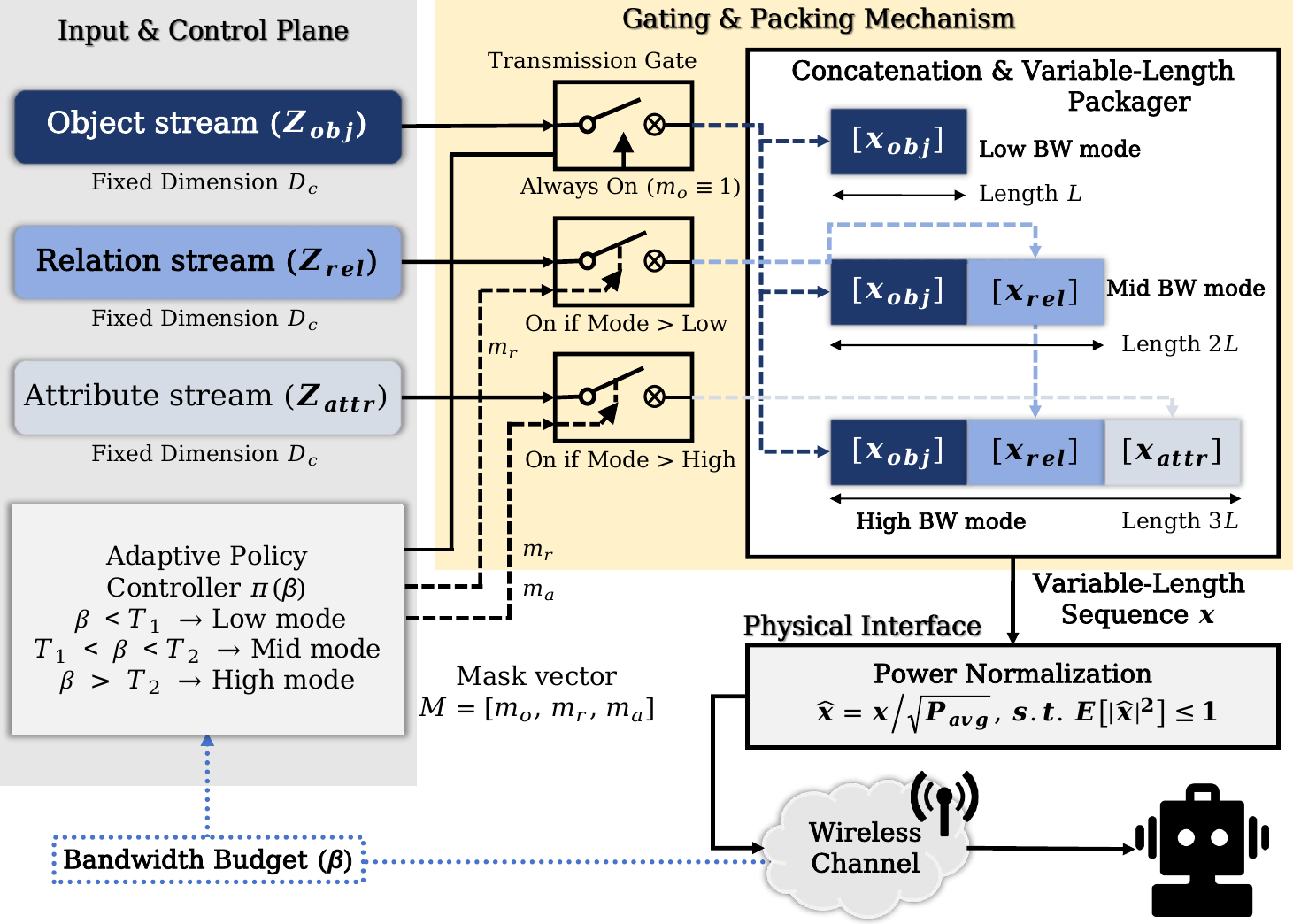}
\caption{\textbf{Block diagram of the bandwidth-adaptive hierarchical transmission.} Governed by a policy controller $\pi(\beta)$, the system dynamically computes an optimal transmission mask $\mathbf{M}$ based on channel bandwidth. This multiplexes prioritized semantic streams (Object, Relation, Attribute) into a variable-length sequence $x$, ensuring robust preservation of core survival semantics under deep fading.}
\label{fig_3}
\end{figure}

To address time-varying wireless channels and bandwidth fluctuations, we design an adaptive hierarchical transmission mechanism based on semantic priority (Fig.~\ref{fig_3}). Unlike traditional pixel-equal video coding paradigms that suffer from indiscriminate performance collapse, our mechanism effectively implements Unequal Error Protection (UEP). By partitioning semantic features into strict priority levels based on the intrinsic logic of the O-A-R graph, the system guarantees intelligent ``graceful degradation'' under severely limited channel resources.

\subsubsection{Resource Allocation and Bandwidth-Adaptive Strategy}

We formulate the hierarchical transmission as a constrained rate-distortion optimization problem. The goal is to maximize the received semantic utility while strictly adhering to the instantaneous bandwidth budget $\beta$. Let $U_\tau$ denote the marginal semantic utility (i.e., importance to the downstream decision task) of stream $\tau \in \{obj, rel, attr\}$, and let $R_\tau$ be its corresponding transmission cost (sequence length). The optimization problem is mathematically defined as:
\begin{equation}
\max_{\mathbf{M}} \sum_{\tau} M_\tau U_\tau \quad \text{s.t.} \quad \sum_{\tau} M_\tau R_\tau \le \beta, \quad M_\tau \in \{0, 1\}
\end{equation}
where $\mathbf{M} = [M_{obj}, M_{rel}, M_{attr}]$ is the transmission mask vector. To solve this, we introduce the Lagrangian formulation:
\begin{equation}
\mathcal{J}(\mathbf{M}, \lambda) = \sum_{\tau} M_\tau U_\tau - \lambda \left( \sum_{\tau} M_\tau R_\tau - \beta \right)
\end{equation}
where $\lambda$ is the Lagrange multiplier dynamically determined by the channel capacity. Because the cognitive hierarchy dictates that $U_{obj} \gg U_{rel} > U_{attr}$, the adaptive policy controller $\pi(\beta)$ yields three progressive operational modes depending on the threshold of $\lambda$:

\begin{itemize}
    \item \textbf{Level 1: Basic Perception Mode (Object-Only, Low BW):} Under severe channel fading where $\beta$ is critically small ($\lambda$ is high), the optimal solution allocates all resources to the object stream ($\mathbf{M} = [1, 0, 0]$). This ``survival mode'' guarantees basic entity identification and obstacle avoidance with merely $1/3$ of the data volume.
    
    \item \textbf{Level 2: Logical Reasoning Mode (Object + Relation, Mid BW):} As channel capacity moderately improves, the relation stream is activated ($\mathbf{M} = [1, 1, 0]$). Since relations define spatial and interaction logic (``cup on table''), this supports complex path planning. Relations are prioritized over attributes due to their higher marginal utility ($U_{rel} > U_{attr}$) in robotic manipulation tasks.
    
    \item \textbf{Level 3: Fine-Grained Operation Mode (Full O-A-R, High BW):} Under broadband channels, the bandwidth constraint relaxes, allowing the complete semantic stream to be transmitted ($\mathbf{M} = [1, 1, 1]$). Fine-grained attributes enable high-precision instructions like ``grasp that red, broken cup.''
\end{itemize}

Consequently, the physical transmission sequence length linearly scales with the available channel state $\beta$, \revb{achieving efficient resource utilization without the all-or-nothing block-outage behavior of Shannon-separated digital schemes}.

\subsubsection{Physical Layer Constraints and Channel Modeling}

To ensure the framework aligns with realistic physical layer conditions, the baseband semantic symbols must satisfy fundamental transmission constraints before being subjected to channel impairments.

\textbf{Average Transmit Power Constraint:} To comply with strict hardware transmission limitations, we implement a statistical normalization layer ensuring constant average power across all bandwidth modes. Any variable-length transmission sequence $\mathbf{x}$ is normalized to a unit power distribution, satisfying the expectation $\mathbb{E}[|\mathbf{x}|^2] \le 1$.

\textbf{AWGN Channel Impairment:} The received baseband signal $\hat{\mathbf{x}}$ over the Additive White Gaussian Noise (AWGN) channel is the superposition of the transmitted symbols and noise: 
\begin{equation}
\hat{\mathbf{x}} = x_{tx} + \mathbf{n}, \quad \mathbf{n} \sim \mathcal{N}(0, \sigma^2 I)
\end{equation}
To maintain mathematically fair comparisons across varying transmission lengths $L$ under different adaptive modes, the noise standard deviation is dynamically scaled such that $\sigma \propto (B \cdot L \cdot \text{SNR})^{-1/2}$. This strictly ensures a consistent channel interference intensity (in dB) for the evaluation of structural fidelity. \rev{Under the block-fading channels additionally evaluated in Sec.~IV, the received signal generalizes to $\hat{\mathbf{x}} = h\,x_{tx} + \mathbf{n}$ with per-block $h\sim\mathcal{CN}(0,1)$ and $\mathbb{E}|h|^2=1$, and the receiver applies perfect-CSI equalization before semantic decoding.}

Furthermore, to optimize the joint source-channel codec for robust operation in time-varying fading environments, the transceiver is jointly optimized over a continuous distribution of channel states (SNR randomly sampled within $[0, 20]$ dB). This forces the semantic decoder to learn highly noise-resilient representations.

\subsection{Semantic Receiver: Cascaded Decoding and Topology Reconstruction}

The semantic receiver performs the inverse communication process, recovering the structured \rev{O-A-R graph} from noisy, truncated channel observations. To resolve dimension inconsistencies from adaptive transmission and exploit O-A-R structural dependencies, the receiver executes three sequential stages (Algorithm~\ref{alg:receiver_decoding}): signal demultiplexing, cascaded recovery, and graph assembly.

\begin{algorithm}[!ht]
\DontPrintSemicolon
\KwIn{Received signal $\mathbf{y}$, Bandwidth mode $\beta$, Mask vector $\mathbf{M}(\beta)$, Threshold $\theta$.}
\KwOut{\rev{O-A-R graph} $G = (\mathcal{V}, \mathcal{E})$.}

    \tcp{1. Demultiplexing \& Error Concealment}
    \For{$\tau \in \{obj, attr, rel\}$}{
        $\hat{\mathbf{x}}_\tau \leftarrow$ extract from $\mathbf{y}$ if $M_\tau(\beta) == 1$, else $\mathbf{0}$\;
        $\hat{\mathbf{z}}_\tau \leftarrow \text{SemanticDecoder}_\tau(\hat{\mathbf{x}}_\tau)$\;
    }
    
    \tcp{2. Cascaded Semantic Recovery}
    $logits_{obj}, prob_{obj}, \mathbf{h}_{entity} \leftarrow \text{EntityDecoder}(\hat{\mathbf{z}}_{obj})$\;
    $\mathcal{P}_{obj} \leftarrow \{i \mid prob_{obj}[i] > \theta\}$\;
    
    $logits_{attr} \leftarrow \text{AttributeEstimator}(\hat{\mathbf{z}}_{attr} \mid \mathbf{h}_{entity})$\;
    
    $\mathcal{E}_{valid} \leftarrow \{(u, v) \mid u, v \in \mathcal{P}_{obj}, u \neq v\}$\;
    $logits_{rel} \leftarrow \text{TopologyEstimator}(\hat{\mathbf{z}}_{rel} \mid \mathbf{h}_{entity})$ on $\mathcal{E}_{valid}$\;
    
    \tcp{3. Graph Assembly}
    $\mathcal{V} \leftarrow$ Assign categories/states to $\mathcal{P}_{obj}$ via $\arg\max(logits_{obj, attr})$\;
    $\mathcal{E} \leftarrow$ Assign directed relations to $\mathcal{E}_{valid}$ via $\arg\max(logits_{rel})$\;
    \Return $G = (\mathcal{V}, \mathcal{E})$\;
\caption{Cascaded Semantic Decoding and Graph Reconstruction}
\label{alg:receiver_decoding}
\end{algorithm}

\textbf{1) Signal Demultiplexing and Error Concealment:} Guided by the negotiated mode $\beta$, the receiver demultiplexes the channel signal into [Object, Relation, Attribute] baseband blocks. To handle intentionally truncated streams under low bandwidths, we employ a \textit{Zero-Padding Error Concealment} strategy. Missing symbols are padded with null vectors (i.e., $\hat{\mathbf{x}}_{\tau} = \mathbf{x}_{\tau}^{rx}$ if $M_{\tau}=1$, else $\mathbf{0}$). This maintains tensor dimensional consistency for the shared decoder back-end and provides an ``information-free'' baseband prior, preventing cascaded estimators from generating spurious false-alarms from random channel noise.

\textbf{2) Cascaded Semantic Signal Recovery:} Recognizing that entity demodulation is a prerequisite for estimating attributes or interactions, we utilize a two-stage cascaded architecture to mitigate error propagation. \textit{Stage 1 (Base Anchor Demodulation):} The EntityDecoder processes the highly-protected $\hat{\mathbf{z}}_{obj}$ stream, yielding detection probabilities and refined baseband vectors $\mathbf{h}_{entity}$. \textit{Stage 2 (Side-Information Aided Estimation):} The reconstructed $\mathbf{h}_{entity}$ is injected as contextual side information for downstream branches, providing strict category constraints for attribute recovery. For topological relations, we apply \textit{Connectivity Search Space Pruning}, evaluating edges exclusively between valid object pairs exceeding a confidence threshold $\theta$. This filters out background noise, yielding the structural distribution $y_{rel} = \mathrm{TopologyEstimator}(\hat{\mathbf{z}}_{rel} \mid \mathbf{h}_{entity})$.

\textbf{3) Task-Oriented Topological Reconstruction:} Ultimately, the receiver bypasses pixel-level reconstruction, outputting a lightweight O-A-R Graph $G=(\mathcal{V}, \mathcal{E})$ for downstream embodied planners. This approach completely decouples semantic perception from physical transmission, ensuring agents maintain essential obstacle avoidance in deep-fading networks (relying solely on Level 1 Object nodes) while executing fine-grained manipulations under high channel capacities.

\subsection{End-to-End System Optimization and JSCC Training}

To minimize the end-to-end semantic transmission errors and maximize the structural fidelity of the recovered topology, we formulate a joint rate-distortion optimization objective. The transceiver is trained via a phased strategy featuring a progressive noise injection mechanism to overcome the inherent gradient instability (cold-start problem) of Joint Source-Channel Coding (JSCC) over highly stochastic channels.

\subsubsection{Semantic Distortion Optimization Objective}

Unlike traditional Shannon-separated systems that optimize exclusively for pixel-level Mean Squared Error (MSE), our joint source-channel codec is optimized to directly minimize the end-to-end semantic distortion $\mathcal{D}_{total}$:
\begin{equation}
\begin{split}
\mathcal{D}_{total} = &\lambda_{obj}\mathcal{D}_{obj} + \lambda_{attr}\mathcal{D}_{attr} + \lambda_{rel}\mathcal{D}_{rel} \\
&+ \lambda_{align}\mathcal{D}_{align} + \lambda_{aux}\mathcal{D}_{aux}
\end{split}
\end{equation}

\textbf{Hierarchical Semantic Distortion ($\mathcal{D}_{obj}, \mathcal{D}_{attr}, \mathcal{D}_{rel}$):} To combat the extreme sparsity and long-tail distribution inherent in topological scene representations, we employ dynamically weighted cross-entropy penalties \cite{lin2017focal}. These metrics strictly penalize the demodulation errors of the entity anchors, discrete states, and logical interactions, respectively.

\textbf{Baseband Signal Alignment ($\mathcal{D}_{align}$):} To foster a compact and noise-resilient latent symbol space, we constrain the JSCC codec using a bandwidth-adaptive combination of Cosine Similarity and L1 regularization:
\begin{equation}
\mathcal{D}_{align} = 10 \cdot (1 - \mathrm{CosSim}(\hat{\mathbf{z}}, \mathbf{z})) + \|\hat{\mathbf{z}} - \mathbf{z}\|_1
\end{equation}
Crucially, only the semantic streams actively transmitted under the current bandwidth mode $\beta$ contribute to this distortion gradient, ensuring dynamic alignment without penalizing truncated signal paths.

\subsubsection{Robust JSCC Optimization Strategy}

Directly optimizing a joint source-channel transceiver from scratch over an AWGN channel often collapses due to the initial random channel noise overwhelming the baseband representations. To guarantee stable convergence, we structure the training pipeline as follows:

\textbf{(1) SNR-Annealed Signal Bypass:} To resolve the initial gradient ambiguity, we introduce a progressive noise injection mechanism (Signal Bypass) that linearly interpolates the noiseless fused baseband signal $\mathbf{z}_{fusion}$ and the noisy channel-corrupted signal $\hat{\mathbf{z}}_{jscc}$:
\begin{equation}
\mathbf{h}_{decoder} = \alpha \cdot \mathbf{z}_{fusion} + (1 - \alpha) \cdot \hat{\mathbf{z}}_{jscc}
\end{equation}
where the decay factor $\alpha$ decreases linearly from 1.0 to 0.0 across the training cycles. This mechanism gradually exposes the cascaded receivers to severe channel noise, smoothing the transition from clean source pre-training to robust noisy inference.

\textbf{(2) Two-Stage Joint Optimization:} We first execute a \textit{Transceiver Warmup} (Stage 1) by isolating the joint source-channel codec to strictly minimize the baseband alignment distortion $\mathcal{D}_{align}$. Subsequently, we perform a \textit{Full End-to-End Joint Optimization} (Stage 2) under a continuous mixed-SNR distribution of $[0, 20]$ dB, accelerated by auxiliary signal supervision ($\mathcal{D}_{aux}$) \cite{lee2015deeply}.

\section{Experiments}
This section systematically verifies the effectiveness of the proposed multimodal O-A-R hierarchical semantic communication framework through extensive experiments. The evaluation demonstrates the framework's capability to overcome semantic transmission bottlenecks of traditional codecs under extremely low bandwidths, its robust graceful degradation against severe channel noise, the crucial semantic compensation from cross-modal synergy, and the alignment of adaptive hierarchical transmission with machine cognitive priority. We also assess reductions in computational complexity and end-to-end latency for real-time embodied intelligence applications.

\subsection{Experimental Setup}
\subsubsection{Dataset Construction}
To evaluate the proposed system under multi-source physical environments, we construct a multimodal perception dataset derived from the Visual Genome (VG) database \cite{krishna2017visual}. The evaluation environment simulates authentic sensory streams of an embodied agent by collecting aligned visual snapshots, descriptive textual instructions, and ambient acoustic signatures.

The dataset comprises 77,855 aligned multimodal samples, split into training (62,565, 80.3\%), validation (7,633, 9.8\%), and test (7,657, 9.8\%) sets. Each sample contains: (i) an RGB image resized to $256 \times 256$ pixels with ImageNet normalization; (ii) a natural language scene description; and (iii) an ambient audio clip sampled at 16\,kHz (max 10\,s). The unified vocabulary consists of 180 visual entity categories, 22 audio event categories (\emph{e.g.}, \textit{footsteps}, \textit{rain}, \textit{siren}), 49 predicate categories, and 95 attribute categories. All annotations are stored in HDF5 format, totaling approximately 328\,MB. Visual relationships account for 94.7\% (avg. 10.4 per sample), while audio event relationships constitute 5.3\% (avg. 0.6 per sample). \rev{The O-A-R graph is built over this closed vocabulary; its relation layer uses the $49$ predicate categories, i.e., $48$ content predicates plus a non-relational \emph{in-image} placeholder.}

\subsubsection{Comparison Baselines}
We compare our method with three representative SOTA baselines across digital and semantic transmission paradigms. For fair comparison, all baselines employ a ``reconstruct-then-evaluate'' pipeline: the receiver reconstructs baseband multimodal data and feeds it into a frozen, unified semantic evaluator to generate the final \rev{O-A-R graph}. This shared back-end ensures performance discrepancies stem solely from the physical transmission schemes.

\begin{figure*}[!t]
\centering
\includegraphics[width=\textwidth]{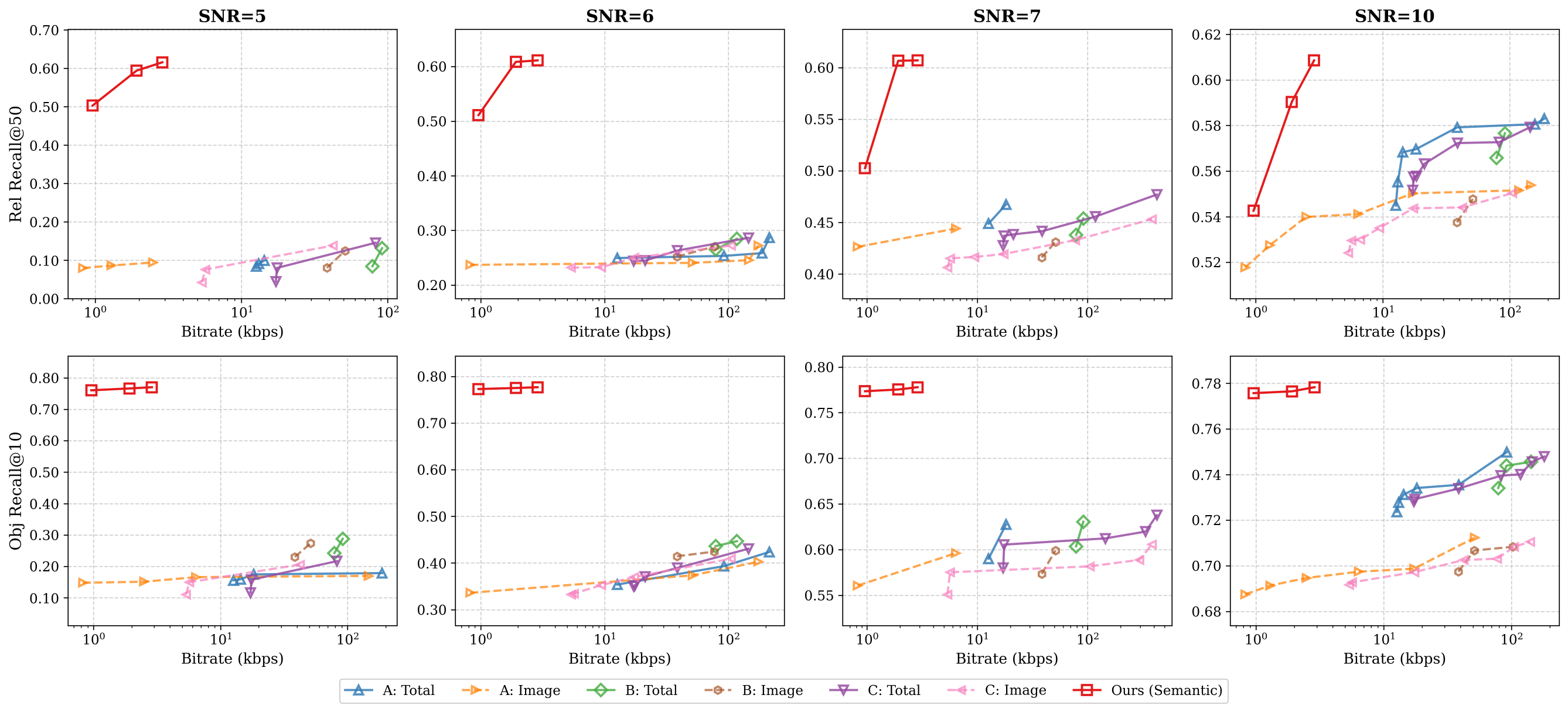}
\caption{\textbf{Comprehensive rate-semantic performance under varying SNR conditions.} Relation Recall@50 (top row) and Object Recall@10 (bottom row) versus total and image-specific data rates (kbps). Our proposed framework maintains high semantic fidelity in the extreme low-bandwidth regime (1--3 kbps), significantly outperforming digital baselines and DeepJSCC, which suffer severe performance drops.}
\label{fig_bps_all}
\end{figure*}

\textbf{Baseline A (BPG + LDPC):} A traditional digital separate scheme. The visual modality is encoded via BPG (based on HEVC/H.265 \cite{sullivan2012overview} Intra), supporting extreme down-sampling for low bitrates. Audio is compressed using Opus \cite{valin2012definition} (6 kbps, VoIP mode), and text via Brotli \cite{alakuijala2018brotli}. Compressed bitstreams are protected by 5G standard LDPC code (rate $1/2$) with 16-QAM modulation.

\textbf{Baseline B (DeepJSCC) \cite{bourtsoulatze2019deep}:} A representative JSCC scheme utilizing an end-to-end convolutional transceiver. The visual source is mapped directly to continuous channel symbols over the AWGN channel, while audio and text rely on standard digital coding (Opus/Brotli + LDPC + 16-QAM).

\textbf{Baseline C (VVC + LDPC):} The next-generation digital separate scheme utilizing VVC/H.266 (VTM Intra) \cite{bross2021overview} for advanced visual compression. In cases of codec incompatibility, it falls back to AVIF/AV1. Audio, text, and channel coding parameters remain identical to Baseline A.

\textbf{Ours (Multimodal Semantic JSCC):} The proposed framework multiplexes multi-source inputs into low-dimensional semantic channel symbols ($N \times D_c$) via hierarchical routing filters. These symbols are transmitted over the un-coded AWGN channel, adaptively scheduled across low (Level 1), mid (Level 2), and high (Level 3) bandwidth modes based on instantaneous channel capacity.

\rev{\textbf{Fairness protocol.} For a fair comparison across paradigms, \revb{all methods are scored on \emph{identical samples, under identical channel realizations, and within a matched bandwidth/CBR budget}, against the same human-annotated ground-truth graphs; the three baselines, which deliver reconstructed sources rather than graphs, are additionally decoded by a \emph{shared frozen semantic evaluator} that converts their reconstructions into the final O-A-R graph (specified below)}. The comparison therefore measures the semantic fidelity attainable per channel use, which is the relevant deployment objective. Under this protocol, the reconstruct-then-predict baselines necessarily spend part of the shared budget recovering pixel detail that is irrelevant to the semantic task; the ``Image'' vs.\ ``Total'' bandwidth curves in Fig.~\ref{fig_bps_all} quantify this overhead. As a paradigm-neutral cross-check, we additionally compare all methods on a downstream zero-shot VQA task under deep fading (Fig.~\ref{fig:vqa_results}), where every method is judged solely by answer accuracy rather than by any reconstruction- or graph-specific metric.}

\revb{\textbf{Evaluator specification and ground truth.} The frozen evaluator is an instance of the same multimodal semantic architecture as our transmitter-side model (modality front-ends, cross-modal fusion, and the cascaded O-A-R prediction heads), trained \emph{channel-free}: it maps clean multimodal inputs directly to O-A-R graphs on the $62{,}565$-sample training split, supervised by the human-annotated Visual Genome graphs, and its weights are frozen thereafter. Four safeguards preclude bias in favor of the proposed method. (i)~The evaluator is trained exclusively on clean source data, before and independently of every transmission experiment; it never observes the reconstructions of any baseline or the channel-corrupted features of our scheme. (ii)~All reported metrics measure agreement with the human-annotated ground-truth graphs, which are \emph{never} produced by the evaluator---the evaluator cannot define the target it is scored against. (iii)~The evaluator serves \emph{only} the baselines, and its clean training distribution matches their output format: the more faithful a baseline's reconstruction, the closer its input lies to the evaluator's training distribution, so this design favors, rather than penalizes, the reconstruct-then-predict paradigm; the clean-input accuracy of the channel-free semantic model (e.g., Object Recall@10 $=83.2\%$, the full-modality row of Table~\ref{tab:ablation_multimodal}) accordingly upper-bounds what any baseline can attain through the evaluator as bandwidth and SNR grow. (iv)~The zero-shot VQA task above provides a paradigm-neutral cross-check, judged solely by answer accuracy by an external vision--language model. Finally, we clarify that our own method involves no additional evaluator: the proposed receiver's cascaded decoder outputs the O-A-R graph directly (Fig.~\ref{fig_2}), and it is this direct output that is scored; the frozen evaluator exists solely to grant the reconstruction-based baselines a graph-producing back-end.}

\subsubsection{Channel Conditions and Evaluation Metrics}
The physical layer transmission is simulated over Additive White Gaussian Noise (AWGN) channels. To emulate realistic physical layer protocols, the digital baselines (A and C) transmit compressed bitstreams via segmented transport blocks rather than a monolithic payload. Their channel transmission is modeled via Shannon capacity simulation with an additional 1.5 dB penalty. Specifically, a decoding outage is triggered if any individual transport block fails (i.e., when the transmission rate strictly exceeds the penalized channel capacity), reflecting the strict data integrity prerequisites of modern predictive video codecs. \rev{To reflect the dynamic channels emphasized for embodied agents, beyond AWGN we additionally evaluate block-fading conditions---Rayleigh and Rician with $K\in\{2,8\}$---using per-block coefficients $h\sim\mathcal{CN}(0,1)$ with $\mathbb{E}|h|^2=1$ and receiver-side perfect-CSI equalization; under fading, the digital baselines additionally suffer per-block outage whenever the instantaneous SNR (scaled by $|h|^2$) falls below their decoding threshold. The corresponding robustness results are reported in the fading evaluation below.}

\begin{table*}[!t]
\centering
\caption{Semantic delivery fidelity and comprehensive accuracy at extreme low bandwidth (SNR = 6 dB). \revb{In each column, \textbf{bold} $=$ best and \underline{underline} $=$ second best.}}
\label{tab:semantic_fidelity_snr6}
\resizebox{\textwidth}{!}{
\begin{tabular}{lcccccccc}
\toprule
\multirow{2}{*}{Method} & \multicolumn{3}{c}{Object Det. (R/P)} & \multicolumn{4}{c}{Relation (R/mR)} & \multirow{2}{*}{GED $\downarrow$} \\
\cmidrule(lr){2-4} \cmidrule(lr){5-8}
 & R/P@5 & R/P@10 & R/P@20 & R/mR@10 & R/mR@15 & R/mR@20 & R/mR@50 & \\
\midrule
Baseline A (BPG, $\sim$13.1 kbps) & 24.3 / 47.5 & 35.2 / 36.1 & 40.7 / 21.4 & 11.5 / 9.5 & 14.4 / 12.1 & 16.9 / 15.4 & 23.0 / 20.8 & 1.16 \\
Baseline B (DeepJSCC, $\sim$78.7 kbps) & 27.2 / 55.3 & 43.6 / 45.9 & 52.6 / 28.2 & 13.7 / 11.0 & 17.1 / 13.9 & 20.0 / 17.5 & 26.5 / 22.1 & 1.05 \\
Baseline C (VVC, $\sim$17.5 kbps) & 23.7 / 48.0 & 35.2 / 36.9 & 42.3 / 22.6 & 12.1 / 9.6 & 15.2 / 11.5 & 18.3 / 14.9 & 23.2 / 18.5 & 1.15 \\
\midrule
Ours (Low BW, 0.96 kbps) & 46.9 / \underline{95.5} & 77.3 / 82.0 & 90.7 / 49.5 & 24.7 / 12.8 & 33.5 / 15.7 & 39.4 / 20.0 & 51.1 / 27.2 & \underline{0.70} \\
Ours (Mid BW, 1.92 kbps) & \underline{47.0} / \textbf{95.9} & \underline{77.5} / \underline{82.8} & \underline{90.8} / \underline{49.6} & \underline{28.6} / \underline{19.5} & \underline{36.9} / \underline{24.1} & \underline{42.3} / \underline{29.7} & \underline{60.9} / \underline{52.5} & \textbf{0.68} \\
Ours (High BW, 2.88 kbps) & \textbf{47.1} / \textbf{95.9} & \textbf{77.7} / \textbf{83.0} & \textbf{90.9} / \textbf{49.7} & \textbf{29.4} / \textbf{19.8} & \textbf{37.0} / \textbf{26.1} & \textbf{43.2} / \textbf{31.7} & \textbf{61.1} / \textbf{52.7} & \textbf{0.68} \\
\bottomrule
\end{tabular}
}
\end{table*}

Regarding evaluation metrics, Semantic Recall@K is adopted to measure the structural fidelity of the recovered \rev{O-A-R graph}, while Graph Edit Distance (GED)\cite{sanfeliu2012distance} is utilized to assess the global topological error. The communication overhead is strictly quantified by the effective data rate (bits per sample) and the Channel Bandwidth Ratio (CBR). {\color{black}We define the evaluation metrics formally as follows. Let the ground-truth graph contain a set $\mathcal{T}$ of semantic items---object nodes (class labels) and relation triplets $(s,p,o)$ denoting (subject class, predicate, object class). Predicted items are ranked by confidence, and \textbf{Recall@K} is the fraction of ground-truth items recovered within the top-$K$ predictions:
\begin{equation}
\mathrm{R}@K=\frac{1}{|\mathcal{T}|}\sum_{t\in\mathcal{T}}\mathbb{1}\!\left[t\in\mathrm{Top}_K\right].
\end{equation}
\textbf{Object Recall@K} and \textbf{Relation Recall@K} instantiate this over object nodes and relation triplets, respectively; \revb{\textbf{Precision@K}, reported alongside recall as the R/P pair in the result tables, is the fraction of the top-$K$ predictions that match a ground-truth item;} \textbf{mean Recall} (mR@K) averages the per-predicate recall to counter class imbalance, and \textbf{Semantic Recall@K} is the overall item-level Recall@K over the O-A-R graph. The \textbf{Graph Edit Distance} (GED) captures global topological error as the minimum total cost of node/edge edits---insertions, deletions, and substitutions---that transform the predicted graph $G_p$ into the ground-truth graph $G_t$, normalized by graph size:
\begin{equation}
\mathrm{GED}(G_p,G_t)=\frac{1}{|G_t|}\min_{(e_1,\dots,e_m)\in\mathcal{P}(G_p,G_t)}\sum_{i=1}^{m}c(e_i),
\end{equation}
where $\mathcal{P}(G_p,G_t)$ denotes the set of edit paths and $c(\cdot)$ the unit edit cost. Finally, the \textbf{Channel Bandwidth Ratio} (CBR) quantifies communication overhead as
\begin{equation}
\mathrm{CBR}=\frac{k}{n},
\end{equation}
where $k$ is the number of transmitted channel symbols and $n$ the source dimension; the effective data rate (bits per sample) is reported alongside.}

\begin{table}[htbp]
\centering
\caption{Minimum bandwidth required to achieve target semantic accuracy across different channel conditions.}
\label{tab:bandwidth_saving}
\setlength{\tabcolsep}{4pt}
\begin{tabular}{l c cc c}
\toprule
\multirow{2}{*}{SNR} & Target & \multicolumn{2}{c}{Min. Bitrate (kbps) $\downarrow$} & \textbf{BW} \\
\cmidrule(lr){3-4}
 & Rel R@50 & Best Baseline & Ours & \textbf{Saving} \\
\midrule
\textbf{5 dB} & $\ge$ 20\% & Failed* & 0.96 (Achieved 50.3\%) & N/A \\
\textbf{6 dB} & $\ge$ 25\% & 38.74 (VVC) & 0.96 (Achieved 51.1\%)& $>$ 97.5\% \\
\textbf{7 dB} & $\ge$ 40\% & 12.67 (BPG) & 0.96 (Achieved 50.2\%)  & $>$ 92.4\% \\
\textbf{10 dB} & $\ge$ 50\% & 12.67 (BPG) & 0.96 (Achieved 54.2\%) & $>$ 92.4\% \\
\bottomrule
\multicolumn{5}{p{0.95\columnwidth}}{\footnotesize \textit{* \revb{At SNR $=5$\,dB the} digital baselines failed to reach the target accuracy at any tested bitrate (Max achieved: 14.6\%). Conversely, our method consistently achieved over 50\% Rel R@50 at just 0.96 kbps across all SNRs.}}
\end{tabular}
\end{table}

\subsection{Rate-Semantic Performance at Extreme Low Bandwidth}
We first evaluate the rate-semantic performance of the proposed system under various bandwidth constraints. Fig.~\ref{fig_bps_all} presents the Semantic Recall versus communication overhead (measured in kbps) across different signal-to-noise ratios (SNR = 5, 6, 7, and 10 dB). This visualization simultaneously captures both the isolated visual transmission cost (dashed lines, ``Image'') and the total multimodal bandwidth overhead (solid lines, ``Total'') of comparative baselines against our method. Quantitative assessments of structural fidelity and bandwidth reduction are detailed in Table~\ref{tab:bandwidth_saving} and Table~\ref{tab:semantic_fidelity_snr6}. The results across all configurations demonstrate that our method \rev{achieves a substantial advantage in the extremely low bandwidth region}. \revb{At SNR $=6$\,dB the low-bandwidth tier alone attains $1.8$--$2.2\times$ the Object Recall@10 and Relation Recall@50 of every baseline}, \rev{at roughly an order of magnitude lower bandwidth (a $>\!90\%$ saving at 1--3 kbps)}.

\textbf{Superior Semantic Density:} As depicted in the leftmost regions of Fig.~\ref{fig_bps_all}, our O-A-R semantic communication scheme operates efficiently at extremely low data rates (typically 1 to 3 kbps). While traditional schemes completely fail to establish meaningful communication in this regime, our method consistently maintains a high Object Recall (above 0.7) and Relation Recall (above 0.4). By directly transmitting highly compressed O-A-R semantic features, our method completely bypasses pixel reconstruction redundancy, preserving remarkably high semantic density.

\textbf{Structural Fidelity and Comprehensive Accuracy:} Beyond raw recall curves, evaluating structural fidelity is crucial. Table~\ref{tab:semantic_fidelity_snr6} details the \revb{Recall/Precision (R/P)}, mean Recall (mRecall), and Graph Edit Distance (GED) at extreme low bandwidth under SNR = 6 dB. While baseline methods require significantly higher bitrates (over 12 kbps) just for basic functionality, they still suffer from high GED and poor mRecall. In contrast, our O-A-R representation operates at a fraction of the bandwidth (0.96--2.88 kbps) while achieving significantly lower GED and higher mRecall. This proves that our approach robustly preserves long-tail semantic relations and global graph topology, which traditional codecs and DeepJSCC severely corrupt at high compression ratios.

\begin{figure*}[!t]
\centering
\includegraphics[width=\textwidth]{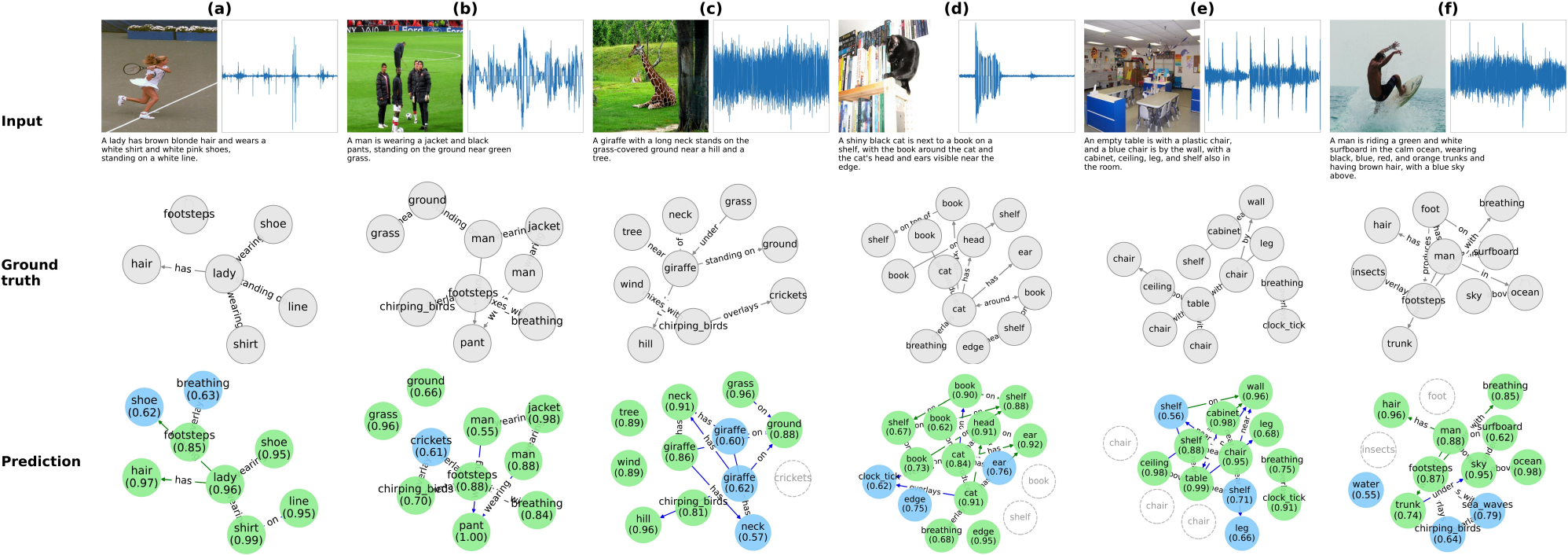}
\caption{\revb{\textbf{Qualitative O-A-R reconstruction} at the low bandwidth tier (1--3 kbps). Rows (top to bottom): multimodal input (image, audio waveform, and text); ground-truth graph (neutral gray); predicted graph. Predicted nodes are drawn in one-to-one positional correspondence with their ground-truth counterparts (green $=$ correctly recovered, blue $=$ spurious), and missed objects appear as dashed empty circles at those positions. Columns (a)--(d) are representative samples; (e,\,f) are failure cases.}}
\label{fig:qualitative}
\end{figure*}

\textbf{Beating SOTA Digital Codecs and Achieving 10-Fold Savings:} Under stringent bitrate constraints, digital separate schemes including Baseline A (BPG) and Baseline C (VVC) suffer from severe quantization artifacts and block effects. As evident in Fig.~\ref{fig_bps_all}, this pixel-level degradation causes the downstream semantic evaluator to fail, leading to abrupt drops in semantic recall. Table~\ref{tab:bandwidth_saving} compares the minimum bitrate required by each method to achieve a target baseline accuracy. At extreme conditions (SNR $\le 5$ dB), traditional schemes completely fail to reach the threshold regardless of allocated bandwidth. Across all functional SNR regimes, our framework guarantees superior semantic delivery while achieving over 90\% bandwidth reduction compared to the most efficient baselines, breaking through the performance bottleneck of Shannon separation systems.

\textbf{Beating DeepJSCC Paradigms:} We further compare against Baseline B (DeepJSCC). Although DeepJSCC mitigates the abrupt ``cliff effect'' of digital quantization to some extent, its joint optimization objective remains implicitly tied to global visual feature reconstruction. Without a mechanism to prioritize semantic importance, DeepJSCC inevitably squanders limited channel capacity on task-irrelevant background noise and texture details. Our framework explicitly decouples the source signal to transmit purely decision-oriented semantics. Both the experimental curves (Fig.~\ref{fig_bps_all}) and quantitative evaluations consistently indicate that under extremely low symbol rates, DeepJSCC experiences steep decline in semantic fidelity, whereas our O-A-R framework robustly retains critical task semantics, maximizing effective information throughput for downstream machine decision-making.

{\color{black}\textbf{Qualitative low-bandwidth reconstruction.} To make concrete what the O-A-R graph preserves versus discards under the extreme budget, Fig.~\ref{fig:qualitative}\revb{(a)--(d)} visualizes representative samples at the low bandwidth tier (1--3 kbps). Even at this budget, the framework preserves the core object anchors and their dominant relations (e.g., the tennis player and her attire, and the soccer players on the field). \revb{Two error patterns remain. Weakly grounded audio events surface as low-confidence spurious nodes, such as \emph{breathing} in (a) and \emph{crickets} in (b). Instances of a repeated class are also duplicated or dropped. The giraffe scene in (c) yields two spurious \emph{giraffe} nodes and a second \emph{neck}, whereas the cluttered scene in (d) misses several same-class \emph{book} and \emph{shelf} instances. Both patterns are examined in the failure-mode analysis below.}}

\begin{table}[htbp]
\centering
\caption{Inference efficiency and complexity (AMD EPYC 9654 CPU, NVIDIA RTX 3090 GPU). \rev{Execution device per stage---A/C: CPU codec, GPU prediction; B: GPU vision JSCC, CPU audio/text; Ours: end-to-end GPU.} \revb{In each latency column, \textbf{bold} $=$ best and \underline{underline} $=$ second best; FLOPs and parameter counts are listed for reference and are not ranked.}}
\label{tab:efficiency_real}
\setlength{\tabcolsep}{3pt} 
\footnotesize 
\begin{tabular}{lccccc}
\toprule
\multirow{2}{*}{\textbf{Method}} & \multicolumn{3}{c}{\textbf{Latency (ms/sample)}} & \multirow{2}{*}{\textbf{FLOPs}} & \multirow{2}{*}{\textbf{Params}} \\
\cmidrule{2-4}
 & \textbf{Tx ($T_{enc}$)} & \textbf{Rx ($T_{dec+pred}$)} & \textbf{Total} & & \\
\midrule
BPG & $219.5 \pm 11.3$ & $230.9 \pm 9.0$ & 450.4 & N/A & 50.93 \\
DeepJSCC & $220.0 \pm 5.8$ & $255.5 \pm 8.7$ & 475.5 & 54.65 & 60.90 \\
VVC/AVIF & $853.8 \pm 25.5$ & $610.1 \pm 14.6$ & 1463.9 & N/A & 50.93 \\
\midrule
\textbf{Ours (Low)} & $80.7 \pm 3.7$ & \underline{$5.8 \pm 0.3$} & 86.4 & 224.77 & 50.93 \\
\textbf{Ours (Mid)} & \underline{$79.7 \pm 2.3$} & $5.9 \pm 0.0$ & \underline{85.7} & 224.74 & 50.93 \\
\textbf{Ours (High)} & $\mathbf{79.2 \pm 1.7}$ & $\mathbf{5.7 \pm 0.3}$ & \textbf{85.0} & 224.74 & 50.93 \\
\bottomrule
\end{tabular}
\end{table}

\textbf{Inference Efficiency and Latency Reduction:} As summarized in Table~\ref{tab:efficiency_real}, we evaluate end-to-end latency and complexity on an AMD EPYC 9654 CPU and NVIDIA RTX 3090 GPU. Digital baselines (A and C) rely on CPU-bound traditional codecs (VVC, BPG), incurring prohibitive encoding delays ($T_{enc} > 200$ ms for A, $> 850$ ms for C) and requiring a two-stage CPU-decoding/GPU-inference pipeline. Baseline B (DeepJSCC) leverages the GPU for vision but bottlenecks on CPU-based audio/text coding. Conversely, our O-A-R framework operates entirely end-to-end on the GPU. By directly transmitting compact semantic vectors and bypassing pixel-level reconstruction, we reduce total latency to $\sim$85 ms---a 5$\times$ to 17$\times$ speedup over baselines---demonstrating exceptional suitability for latency-sensitive, real-time embodied intelligence. \rev{Notably, our model has \emph{higher} nominal FLOPs than DeepJSCC, yet a lower wall-clock latency, because the traditional and DeepJSCC pipelines are bottlenecked by CPU-bound codec and entropy-coding stages. We therefore report both the FLOPs, as a hardware-neutral measure of computational complexity, and the wall-clock latency, which reflects an end-to-end GPU deployment.}

\begin{figure*}[!t]
\centering
\includegraphics[width=\textwidth]{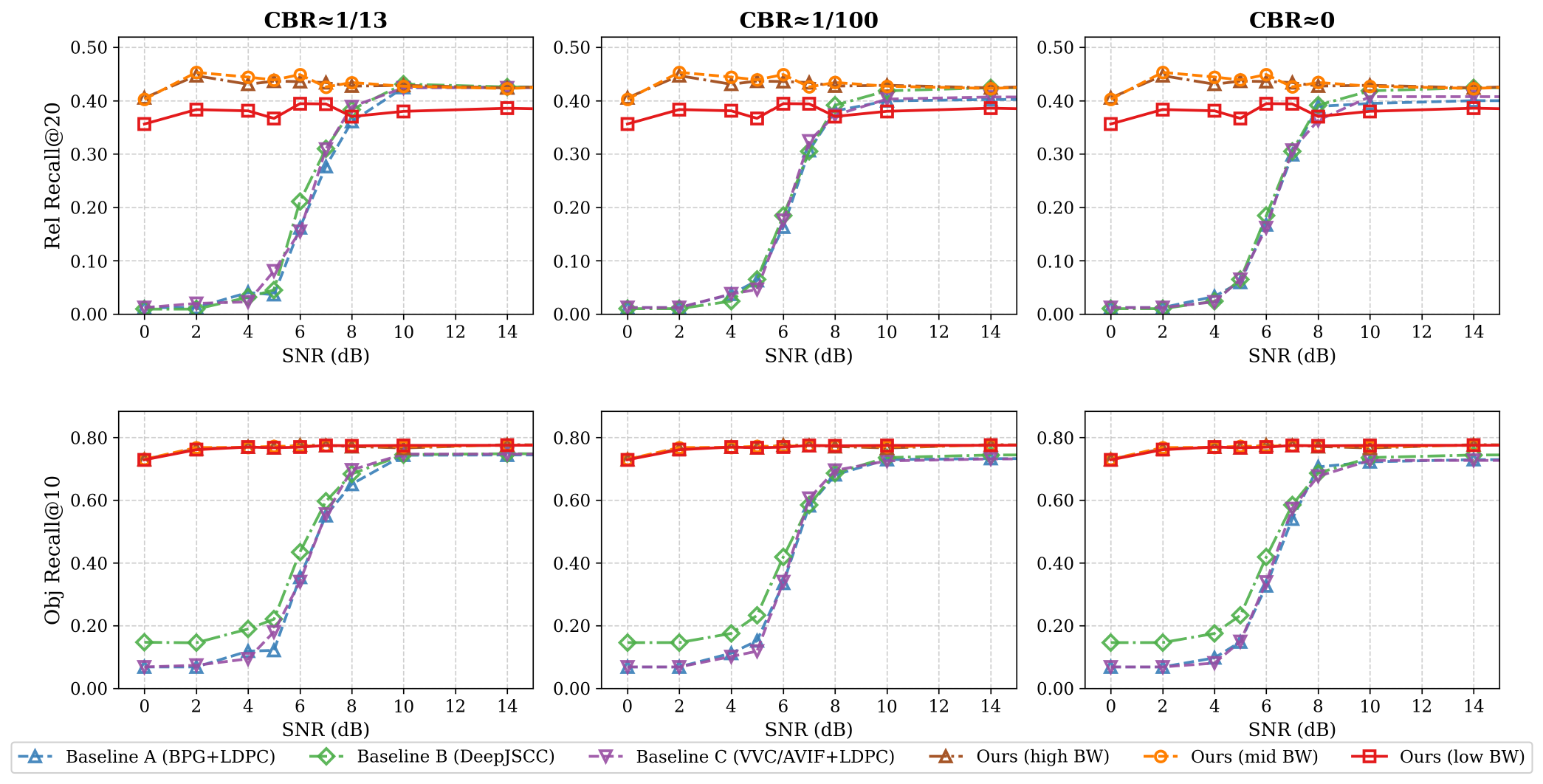}
\caption{\textbf{System robustness and graceful degradation under varying channel conditions.} Relation Recall@20 (top row) and Object Recall@10 (bottom row) versus SNR under varying CBR constraints. While traditional baselines suffer severe ``cliff effects'' below 8 dB, our adaptive mechanism ensures graceful degradation, robustly preserving foundational object semantics even in deep fading (SNR $\le$ 4 dB).}
\label{fig_snr}
\end{figure*}

\begin{table*}[t]
\centering
\caption{Decoding failure rate and Graph Edit Distance (GED) across the critical SNR boundaries.}
\label{tab:cliff_effect}
\begin{tabular}{l ccc ccc ccc ccc}
\toprule
\multirow{2}{*}{Method} & \multirow{2}{*}{CBR} & \multicolumn{2}{c}{SNR = 4 dB} & & \multicolumn{2}{c}{SNR = 6 dB} & & \multicolumn{2}{c}{SNR = 8 dB} & & \multicolumn{2}{c}{SNR = 10 dB} \\
\cmidrule{3-4} \cmidrule{6-7} \cmidrule{9-10} \cmidrule{12-13}
& & Fail Rate $\downarrow$ & GED $\downarrow$ & & Fail Rate $\downarrow$ & GED $\downarrow$ & & Fail Rate $\downarrow$ & GED $\downarrow$ & & Fail Rate $\downarrow$ & GED $\downarrow$ \\
\midrule
Baseline A (BPG) & 1/13 & 100.0\% & 1.41 & & 66.4\% & 1.05 & & 7.8\% & 0.74 & & 0.8\% & 0.68 \\
Baseline C (VVC) & 1/13 & 100.0\% & 1.41 & & 59.4\% & 1.18 & & 15.6\% & 0.87 & & 0.0\% & 0.68 \\
Baseline B (DeepJSCC) & 1/13 & N/A* & 1.31 & & N/A* & 1.05 & & N/A* & 0.77 & & N/A* & 0.70 \\
\midrule
\textbf{Ours (Low BW)} & \revb{$\approx$}0 & 0.0\% & 0.71 & & 0.0\% & 0.70 & & 0.0\% & 0.68 & & 0.0\% & 0.67 \\
\textbf{Ours (Mid BW)} & \revb{$\approx$}0 & 0.0\% & 0.69 & & 0.0\% & 0.68 & & 0.0\% & 0.67 & & 0.0\% & 0.67 \\
\textbf{Ours (High BW)} & \revb{$\approx$}0 & 0.0\% & 0.69 & & 0.0\% & 0.68 & & 0.0\% & 0.67 & & 0.0\% & 0.67 \\
\bottomrule
\multicolumn{13}{p{0.95\textwidth}}{\footnotesize \textit{* DeepJSCC operates in an analog-like manner without explicit digital packet drops; thus, binary failure rates are not applicable, though its GED severely degrades.}}
\end{tabular}
\end{table*}

\subsection{Robustness to Channel Variations and Graceful Degradation}
\begin{figure*}[!t]
\centering
\includegraphics[width=\textwidth]{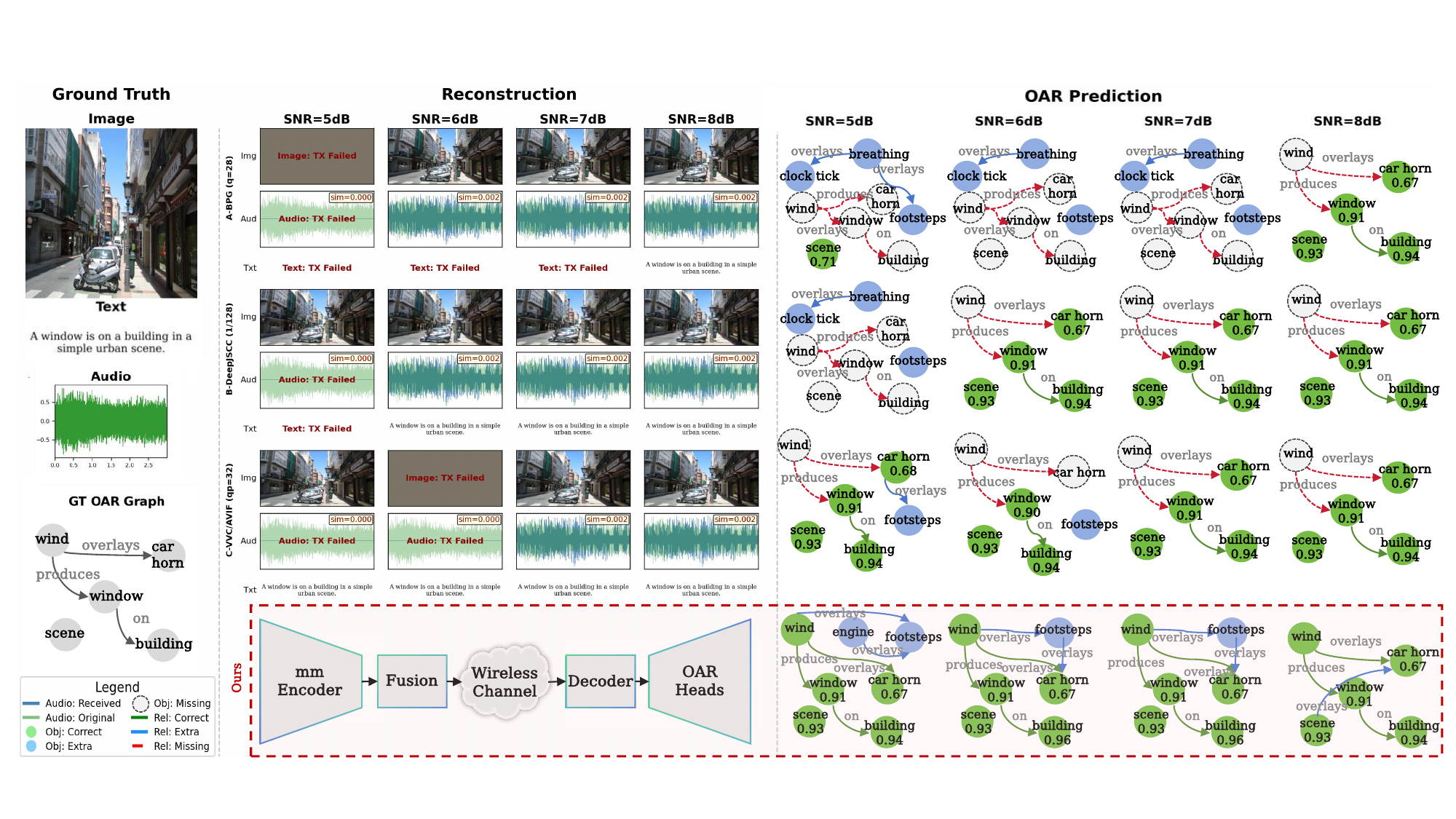}
\caption{\revb{\textbf{Qualitative comparison of multimodal O-A-R graph generation under varying SNRs} (one representative channel realization per panel). Predicted graphs are drawn in one-to-one positional correspondence with the ground-truth graph (leftmost column); node and edge encodings follow the legend.}}
\label{fig:qualitative_snr}
\end{figure*}

To verify noise resistance and the effectiveness of the joint source-channel coding (JSCC) architecture in dynamic wireless environments---such as those characterized by rapid fading and high mobility in embodied robotic applications---we evaluate system performance across a wide continuous spectrum of Signal-to-Noise Ratios (SNR). Fig.~\ref{fig_snr} illustrates Relation Recall@20 and Object Recall@10 under varying SNR levels (0 to 14 dB) subject to strictly enforced Channel Bandwidth Ratio (CBR) constraints (CBR = 1/13, 1/100, and \revb{$\approx$}0). To comprehensively understand the underlying topological integrity and failure mechanisms, we detail decoding outage rates and fine-grained semantic metrics in Table~\ref{tab:cliff_effect} and Table~\ref{tab:graceful_degradation}, alongside qualitative visual baseband recovery comparisons in Fig.~\ref{fig:qualitative_snr}.

\textbf{The Cliff Effect of Traditional Schemes:} As observed in the performance curves and quantified in Table~\ref{tab:cliff_effect}, digital separate communication schemes (Baseline A and C) exhibit a typical ``cliff effect'' when channel quality deteriorates. When the instantaneous SNR drops below a critical threshold (approximately 6--8 dB), the required transmission rate exceeds the Shannon channel capacity. The channel bit error rate then overwhelms LDPC error correction capability, triggering an avalanche of unrecoverable block errors. Since modern codecs rely on highly stateful entropy coding and spatial prediction loops, the loss of even a single transport block breaks decoding context synchronization, rendering the remaining bitstream unparsable.

Table~\ref{tab:cliff_effect} reveals that at SNR = 4 dB, this vulnerability results in a 100\% decoding outage rate. The Graph Edit Distance (GED) explodes to over 1.40, indicating complete topological destruction and causing semantic recall metrics to plummet to near zero. Baseline B (DeepJSCC) similarly experiences sharp decline in this deep-fading region. As shown by its rapidly degrading GED and visually corroborated by severe reconstruction artifacts in Fig.~\ref{fig:qualitative_snr}, its implicit feature reconstruction becomes highly vulnerable to channel noise, completely failing to provide reliable perceptual data in deep fading scenarios.

\begin{table*}[!t]
\centering
\caption{Semantic performance and graceful degradation in hostile channel environments (SNR $\le$ 4 dB). \revb{Within each SNR block, \textbf{bold} $=$ best and \underline{underline} $=$ second best, applied separately to each metric of an R/P (R/mR) pair.}}
\label{tab:graceful_degradation}
\begin{tabular*}{\textwidth}{@{\hspace{1em}\extracolsep{\fill}}ll ccc cccc@{\hspace{1em}}}
\toprule
\multirow{2}{*}{SNR} & \multirow{2}{*}{Method} & \multicolumn{3}{c}{Object Det. (R/P) $\uparrow$} & \multicolumn{4}{c}{Relation (R/mR) $\uparrow$} \\
\cmidrule(lr){3-5} \cmidrule(lr){6-9}
& & R/P@5 & R/P@10 & R/P@20 & R/mR@10 & R/mR@15 & R/mR@20 & R/mR@50 \\
\midrule
\multirow{5}{*}{\textbf{0 dB}} 
 & Baseline C (VVC) & 6.2 / 12.0 & 6.8 / 6.7 & 9.4 / 4.8 & 1.2 / 0.4 & 1.2 / 0.4 & 1.2 / 0.4 & 1.2 / 0.4 \\
 & Baseline B (DeepJSCC) & 8.7 / 17.5 & 14.7 / 14.9 & 20.5 / 10.7 & 0.9 / 0.3 & 0.9 / 0.3 & 0.9 / 0.3 & 0.9 / 0.3 \\
\cmidrule{2-9}
 & \textbf{Ours (Low BW)} & 45.2 / 93.0 & \underline{72.9} / 77.2 & 87.7 / 47.9 & 23.8 / 10.9 & 31.4 / 13.2 & 35.6 / 15.0 & 43.1 / 20.3 \\
 & \textbf{Ours (Mid BW)} & \textbf{45.8} / \textbf{93.9} & \textbf{73.0} / \underline{77.6} & \underline{88.7} / \underline{48.5} & \underline{27.4} / \textbf{19.5} & \textbf{36.3} / \underline{26.9} & \underline{40.3} / \underline{29.6} & \underline{53.0} / \underline{43.3} \\
 & \textbf{Ours (High BW)} & \underline{45.5} / \underline{93.6} & \textbf{73.0} / \textbf{77.7} & \textbf{89.3} / \textbf{48.7} & \textbf{28.4} / \underline{18.7} & \underline{35.5} / \textbf{27.1} & \textbf{40.5} / \textbf{31.7} & \textbf{55.0} / \textbf{44.9} \\
\midrule
\multirow{5}{*}{\textbf{2 dB}} 
 & Baseline C (VVC) & 6.5 / 12.5 & 7.3 / 7.2 & 10.2 / 5.2 & 1.8 / 0.5 & 2.0 / 0.6 & 2.0 / 0.6 & 2.0 / 0.6 \\
 & Baseline B (DeepJSCC) & 8.9 / 17.8 & 14.5 / 14.8 & 20.3 / 10.5 & 0.9 / 0.3 & 0.9 / 0.3 & 0.9 / 0.3 & 0.9 / 0.3 \\
\cmidrule{2-9}
 & \textbf{Ours (Low BW)} & 46.7 / \underline{95.6} & 76.1 / 80.8 & 89.6 / 48.9 & 23.9 / 10.6 & 32.1 / 13.0 & 38.3 / 15.2 & 49.1 / 22.9 \\
 & \textbf{Ours (Mid BW)} & \textbf{46.9} / \textbf{95.8} & \textbf{76.8} / \textbf{81.4} & \underline{89.8} / \underline{49.0} & \textbf{31.1} / \textbf{24.3} & \textbf{38.9} / \textbf{30.5} & \textbf{45.3} / \textbf{35.7} & \textbf{62.0} / \textbf{52.3} \\
 & \textbf{Ours (High BW)} & \underline{46.8} / 95.3 & \underline{76.4} / \underline{81.1} & \textbf{90.5} / \textbf{49.5} & \underline{29.3} / \underline{20.0} & \underline{37.2} / \underline{27.2} & \underline{44.7} / \underline{32.8} & \underline{59.1} / \underline{49.3} \\
\midrule
\multirow{5}{*}{\textbf{4 dB}} 
 & Baseline C (VVC) & 7.6 / 15.0 & 9.4 / 9.6 & 12.4 / 6.5 & 2.1 / 0.9 & 2.2 / 1.1 & 2.3 / 1.1 & 3.2 / 1.4 \\
 & Baseline B (DeepJSCC) & 10.7 / 21.7 & 18.9 / 19.7 & 25.1 / 13.3 & 2.4 / 1.2 & 2.9 / 1.5 & 3.2 / 1.6 & 3.5 / 2.0 \\
\cmidrule{2-9}
 & \textbf{Ours (Low BW)} & \textbf{46.9} / \textbf{95.8} & \textbf{77.0} / \underline{81.6} & \underline{90.5} / \textbf{49.5} & 24.5 / 11.0 & 32.6 / 15.6 & 38.1 / 18.0 & 51.4 / 26.6 \\
 & \textbf{Ours (Mid BW)} & \underline{46.7} / \underline{95.5} & \underline{76.9} / \underline{81.6} & \textbf{90.6} / \textbf{49.5} & \textbf{30.2} / \textbf{20.4} & \underline{37.9} / \textbf{28.6} & \textbf{44.4} / \textbf{32.6} & \textbf{61.4} / \underline{51.1} \\
 & \textbf{Ours (High BW)} & 46.6 / 95.2 & \textbf{77.0} / \textbf{81.7} & 90.4 / \underline{49.4} & \underline{29.3} / \underline{19.8} & \textbf{38.3} / \underline{25.6} & \underline{43.1} / \underline{31.6} & \underline{60.3} / \textbf{51.2} \\
\bottomrule
\end{tabular*}
\end{table*}

\textbf{Graceful Degradation of the Proposed Framework:} In stark contrast, our O-A-R semantic communication system demonstrates remarkable ``graceful degradation,'' fundamentally breaking the all-or-nothing limitation of Shannon-separated systems. Our decoupled semantic symbols are demodulated independently; by mapping continuous semantic features directly to the physical layer, the architecture bypasses strict binary quantization, directly preventing the cascading error propagation that causes decoding outage. As shown in Table~\ref{tab:cliff_effect}, our method maintains a 0\% decoding failure rate and stably low GED across critical SNR boundaries. Table~\ref{tab:graceful_degradation} unpacks performance in extremely hostile channel conditions (SNR $\le$ 4 dB). While Shannon-based schemes collapse into perceptual ``blindness,'' our system adaptively maintains robust object anchors (\revb{Object Recall@10 above $0.72$}) even as high-level semantics (Relation Recall and mRecall) proportionately degrade. This resilience is visualized in Fig.~\ref{fig:qualitative_snr}, \revb{where, in this qualitative illustration, one representative channel realization is shown per panel. All five ground-truth objects are recovered by the proposed framework at each of the four SNRs, whereas no baseline recovers more than four at any SNR. The baselines also collapse abruptly rather than gradually. BPG recovers a single object at $5$\,dB and none at $6$ and $7$\,dB, DeepJSCC recovers none at $5$\,dB, and VVC loses one object at $6$\,dB. Because a single block error corrupts an entire reconstruction, the semantics that survive a separated transmission follow the particular channel realization rather than the SNR alone.}

By dynamically scheduling transmission modes (High, Mid, Low BW) according to instantaneous channel state information (CSI), the system truncates fragile high-level attributes and relations when the channel worsens, prioritizing reliable delivery of underlying object anchors. This is reflected in Table~\ref{tab:graceful_degradation}: \revb{even at SNR $=0$\,dB the Low BW mode preserves object recall on par with the higher tiers ($\mathrm{R}@10=72.9\%$ versus $73.0\%$ at High BW) while transmitting only the object layer, deliberately trading away relation detail (Relation R@50 of $43.1\%$ versus $55.0\%$ at High BW) to concentrate the constrained budget on object anchors}. This inherent Unequal Error Protection (UEP) is achieved purely through adaptive baseband scheduling, bypassing the latency overhead of conventional layered forward error correction (FEC), ensuring embodied agents retain fundamental survival capabilities (navigation and obstacle avoidance) rather than entirely losing perceptual function in degraded networks.

\textbf{Alignment with Machine Decision Priority:} The graceful degradation mechanism directly answers whether our hierarchical design aligns with upper-layer machine cognitive logic. For embodied intelligence, perceiving spatial existence of entities is the primary prerequisite for safety and navigation, followed by understanding their states (Attributes) and complex interactions (Relations). Unlike traditional codecs or monolithic DeepJSCC paradigms---which treat all visual elements with equal priority and suffer catastrophic, indiscriminate semantic loss at low SNRs---our O-A-R stream decoupling strictly guarantees delivery of core survival-critical information first. By ensuring \revb{that entity existence and relative spatial topology are} preserved even when fine-grained interaction logic is lost, this selective transmission logic explicitly mirrors the cognitive priority of machine vision systems, demonstrating that the proposed hierarchical mechanism optimally balances semantic completeness with strictly constrained transmission costs.

\begin{table}[t]
\centering
\caption{Multimodal ablation study \revb{(clean, channel-free)} on the impact of different modality combinations. \revb{In each column, \textbf{bold} $=$ best and \underline{underline} $=$ second best.}}
\label{tab:ablation_multimodal}
\begin{tabular}{l cc c c}
\toprule
\multirow{2}{*}{\textbf{Modality}} & \multicolumn{2}{c}{\textbf{Object Det. (\%)} $\uparrow$} & \textbf{Relation (\%)} $\uparrow$ & \multirow{2}{*}{\textbf{GED} $\downarrow$} \\
\cmidrule(lr){2-3} \cmidrule(lr){4-4}
& R/P@10 & R/P@20 & R/mR@50 & \\
\midrule
Image (I)           & 43.9 / 43.4 & 59.9 / 30.0 & 20.0 / 13.6 & 0.83 \\
Text (T)            & 48.1 / 47.9 & 58.7 / 29.5 & 28.1 / 21.0 & 0.74 \\
Audio (A)           & 34.1 / 33.0 & 43.0 / 21.1 & 13.0 /  9.9 & 1.04 \\
\midrule
I + T & 56.3 / 56.0 & 71.0 / 35.8 & 32.7 / 23.0 & 0.76 \\
I + A & 59.9 / 58.5 & \underline{76.3} / \underline{38.0} & 30.1 / 21.2 & 0.80 \\
T + A & \underline{61.3} / \underline{60.6} & 70.9 / 35.6 & \underline{39.4} / \underline{28.6} & \underline{0.70} \\
\midrule
(I + T + A) & \textbf{83.2} / \textbf{81.6} & \textbf{93.4} / \textbf{47.0} & \textbf{65.6} / \textbf{55.7} & \textbf{0.66} \\
\bottomrule
\multicolumn{5}{p{0.95\columnwidth}}{\footnotesize \textit{* Note: I=Image, T=Text, A=Audio. The full multimodal synergy significantly improves both semantic retention and structural accuracy (lowest GED).}}
\end{tabular}
\end{table}

{\color{black}\subsection{Robustness under Fading Channels}
The evaluation above considers AWGN. Because embodied agents operate in mobile, multipath environments, we additionally evaluate block-fading channels---Rayleigh and Rician with $K\in\{2,8\}$---using per-block coefficients $h\sim\mathcal{CN}(0,1)$ with $\mathbb{E}|h|^2=1$ and receiver-side perfect-CSI equalization. Our models are finetuned and all baselines retrained under each fading channel; under fading, the digital baselines additionally incur per-block outage whenever the instantaneous SNR (scaled by $|h|^2$) falls below their decoding threshold. Fig.~\ref{fig_fading} plots the robustness curves and Table~\ref{tab:fading} reports representative operating points.

\begin{figure*}[!t]
\centering
\color{black}
\includegraphics[width=\textwidth]{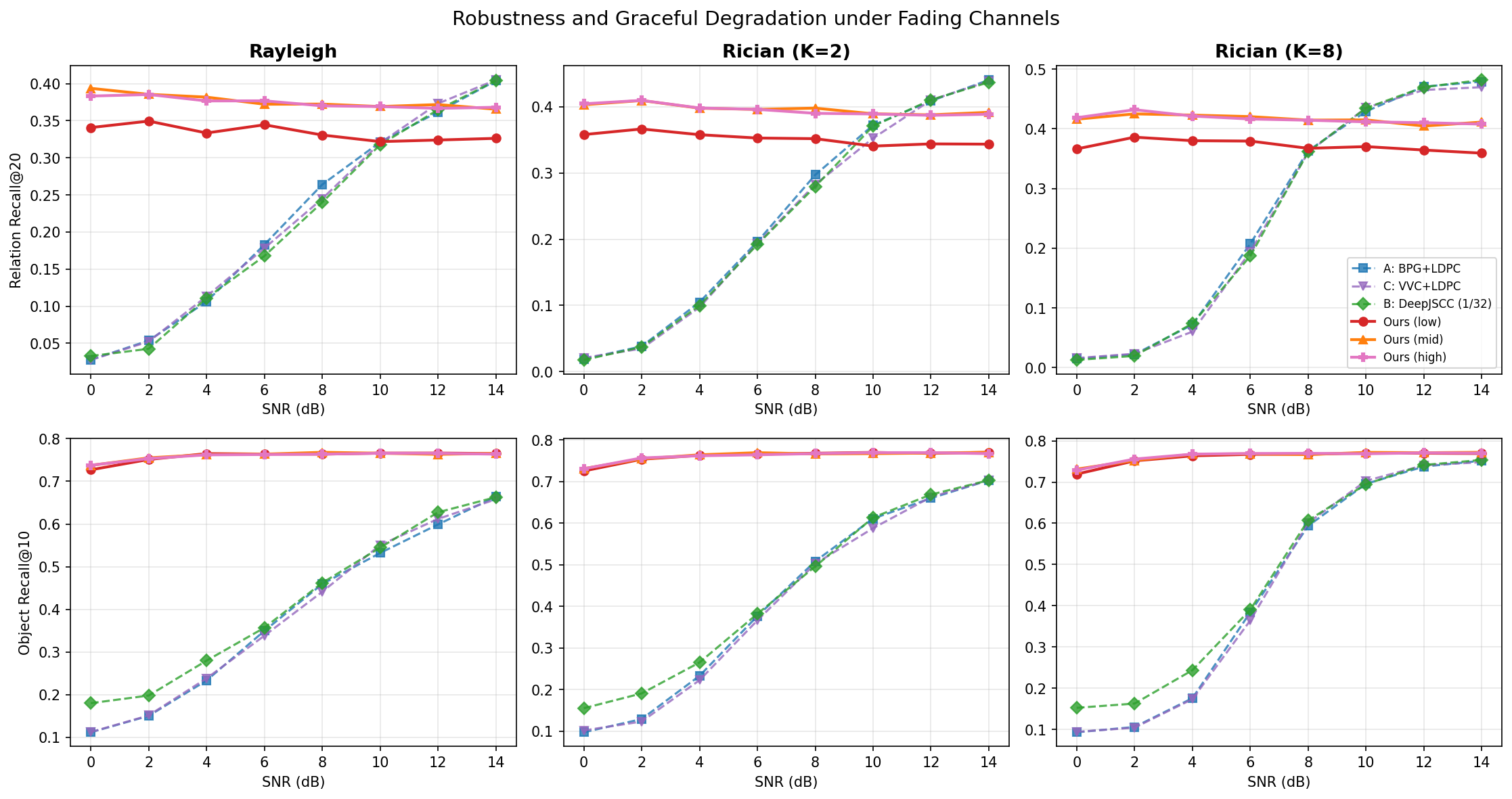}
\caption{\textbf{Robustness under Rayleigh and Rician ($K{=}2,8$) fading.} Relation Recall@20 (top) and Object Recall@10 (bottom) versus SNR. Traditional baselines collapse at low SNR due to per-block outage, whereas our method degrades gracefully and preserves foundational object anchors across all three fading channels.}
\label{fig_fading}
\end{figure*}

\begin{table}[!t]
\centering
\color{black}
\caption{\textbf{Object Recall@10 under fading channels} (mid tier). Baselines at SNR$=0$\,dB; ours at $0$ and $14$\,dB.}
\label{tab:fading}
\small
\begin{tabular}{lccccc}
\toprule
\multirow{2}{*}{Channel} & \multicolumn{3}{c}{Baselines @\,0\,dB} & \multicolumn{2}{c}{Ours (mid)} \\
\cmidrule(lr){2-4}\cmidrule(lr){5-6}
 & A & B & C & @\,0\,dB & @\,14\,dB \\
\midrule
Rayleigh        & 0.112 & 0.162 & 0.112 & 0.641 & 0.796 \\
Rician $K{=}2$  & 0.098 & 0.155 & 0.102 & 0.685 & 0.804 \\
Rician $K{=}8$  & 0.093 & 0.151 & 0.094 & 0.733 & 0.810 \\
\bottomrule
\end{tabular}
\end{table}

As shown in Table~\ref{tab:fading}, all digital and DeepJSCC baselines collapse to an Object Recall@10 of $0.09$--$0.16$ at $0$\,dB across every fading channel, as per-block deep fades trigger decoding outages. In contrast, our framework sustains $0.64$--$0.73$ at $0$\,dB and recovers to $\approx0.80$ at $14$\,dB---a $4$--$7\times$ advantage---degrading gracefully rather than catastrophically. The self-consistent ordering (Rayleigh $<$ Rician $K{=}2$ $<$ Rician $K{=}8$) confirms that performance tracks channel hardness as expected. These results substantiate the graceful-degradation behavior of the proposed framework in genuinely dynamic fading channels.
}

\subsection{Ablation Study on Cross-Modal Semantic Compensation}
To investigate the semantic compensation effect of non-visual modalities (textual instructions and ambient audio) under complex, resource-constrained environments---such as severe visual occlusion, deep channel fading, or sensor failure---we conduct a comprehensive ablation study evaluating different modality combinations. Table~\ref{tab:ablation_multimodal} presents the quantitative comparison in terms of Object Recall@20, Relation Recall@50, and Graph Edit Distance (GED).

\textbf{Superiority of Full Modality Synergy:} The results clearly demonstrate that the Full model integrating all three modalities achieves optimal performance across all metrics, reaching an Object Recall@20 of 93.4\% and a Relation Recall@50 of 65.6\%, while maintaining the lowest structural error with a GED of 0.66. This validates the effectiveness of our cross-modal joint source aggregation module in deeply aligning heterogeneous data into a unified semantic space, successfully mitigating the representational gap between high-dimensional visual grids and sparse non-visual tokens.

\textbf{Cross-Modal Strong Priors:} The ablation on partial modalities further proves the necessity of multimodal compensation. When relying solely on the Image (I) modality, the system only yields an Object Recall@20 of 59.9\% and a Relation Recall@50 of 20.0\% due to visual limitations or compression degradation. Incorporating non-visual priors significantly boosts perception: the Image+Text (I+T) and Image+Audio (I+A) combinations elevate Object Recall@20 to 71.0\% and 76.3\%, respectively. This illustrates that textual instructions serve as a powerful attention guide to resolve visual ambiguities, while environmental audio cues (engine noises or collision sounds) act as crucial supplementary evidence for object localization and classification, compensating for the camera's restricted field of view.

\textbf{Robustness in Extreme Deprivation:} Notably, even when visual input is completely deprived or catastrophically corrupted, the Text+Audio (T+A) configuration can still reconstruct a viable O-A-R Graph, achieving an Object Recall@20 of 70.9\% and a Relation Recall@50 of 39.4\%. Furthermore, its structural error (GED = 0.70) is even lower than those of dual-modalities containing images (I+T and I+A). This counter-intuitive result indicates that under severe fading, highly corrupted visual features may introduce topological hallucinations, whereas lightweight, reliable text and audio streams provide deterministic semantic anchors. This confirms that by eliminating mutual information across modalities, the proposed framework achieves a robustness gain of ``$1+1+1 > 3$'', ensuring the embodied agent's resilient perception and topological reasoning in hostile environments.

{\color{black}\subsection{Hyperparameter Ablation}
To justify the adopted configuration, we ablate the four key design choices---the semantic priority order, the query count $N$, the compression dimension $D_c$, and the relation loss weight $\lambda_{\mathrm{rel}}$---varying one factor at a time while fixing the others. Results are summarized in Table~\ref{tab:hyperparam_ablation}.

\begin{table}[!t]
\centering
\color{black}
\caption{\textbf{Hyperparameter ablation.} Each group varies one factor (priority at the mid tier, $D_c$ and $\lambda_{\mathrm{rel}}$ at the high tier, \revb{all channel groups at SNR $=20$\,dB}; $N$ channel-free). \revb{Within each group, \textbf{bold} $=$ best and \underline{underline} $=$ second best; $^{\dagger}$ marks the adopted setting.} $\uparrow/\downarrow$: higher/lower better.}
\label{tab:hyperparam_ablation}
\small
\begin{tabular}{lcccc}
\toprule
Setting & Obj R@10\,$\uparrow$ & Rel R@20\,$\uparrow$ & Rel R@50\,$\uparrow$ & GED\,$\downarrow$ \\
\midrule
\multicolumn{5}{l}{\textit{Priority order (mid tier)}}\\
O$>$A$>$R & \revb{\underline{0.808}} & \revb{\underline{0.316}} & \revb{\underline{0.456}} & \revb{\textbf{0.581}} \\
O$>$R$>$A\revb{$^{\dagger}$} & \revb{\textbf{0.824}} & \revb{\textbf{0.475}} & \revb{\textbf{0.631}} & \revb{\underline{0.627}} \\
\midrule
\multicolumn{5}{l}{\textit{Query count $N$}}\\
20 & \revb{\underline{0.811}} & \revb{\underline{0.494}} & \revb{\underline{0.631}} & \revb{\underline{0.637}} \\
30\revb{$^{\dagger}$} & \revb{\textbf{0.833}} & \revb{\textbf{0.511}} & \revb{\textbf{0.658}} & \revb{0.659} \\
50 & \revb{0.788} & \revb{0.306} & \revb{0.436} & \revb{\textbf{0.537}} \\
\midrule
\multicolumn{5}{l}{\textit{Compression dimension $D_c$}}\\
16 & \revb{0.746} & \revb{0.280} & \revb{0.407} & \revb{\underline{0.623}} \\
32\revb{$^{\dagger}$} & \revb{\textbf{0.824}} & \revb{\textbf{0.474}} & \revb{\textbf{0.632}} & \revb{0.627} \\
64 & \revb{\underline{0.804}} & \revb{\underline{0.362}} & \revb{\underline{0.511}} & \revb{\textbf{0.575}} \\
\midrule
\multicolumn{5}{l}{\textit{Relation loss weight $\lambda_{\mathrm{rel}}$}}\\
0.2 & \revb{\underline{0.807}} & \revb{0.368} & \revb{\underline{0.520}} & \revb{\underline{0.574}} \\
1.0\revb{$^{\dagger}$} & \revb{\textbf{0.824}} & \revb{\textbf{0.474}} & \revb{\textbf{0.632}} & \revb{0.627} \\
5.0 & \revb{0.806} & \revb{\underline{0.370}} & \revb{\underline{0.520}} & \revb{\textbf{0.570}} \\
\bottomrule
\end{tabular}
\end{table}

\textbf{Priority order.} At an equal two-layer budget, transmitting relations in the middle tier (O$>$R$>$A) raises Relation Recall@20/@50 by \revb{$+50\%$/$+38\%$} over O$>$A$>$R while object recall is \revb{essentially unchanged}, confirming that relations carry higher marginal utility than attributes and validating the O$>$R$>$A ordering.

\textbf{Query count.} $N{=}30$ achieves the best object recall---the foundational layer of the graph---together with \revb{the best relation recall}. Increasing to $N{=}50$ yields no benefit and in fact degrades recall (under an equal training budget each additional query is less well trained), while reducing to $N{=}20$ sacrifices object coverage. We therefore adopt $N{=}30$.

\textbf{Compression dimension.} $D_c{=}16$ collapses semantic fidelity (\revb{$-36\%$ in Rel R@50 and $-9\%$ in object recall}), whereas $D_c{=}64$ provides no gain over $D_c{=}32$ at twice the bandwidth cost. $D_c{=}32$ is thus the efficient operating point.

\textbf{Loss weight.} \revb{Relation recall peaks at $\lambda_{\mathrm{rel}}{=}1$ (a clear inverted-U over $[0.2,5]$) while object recall stays essentially constant}; we adopt $\lambda_{\mathrm{rel}}{=}1$.

\revb{\textbf{Selection criterion.} All four choices are made on the recall metrics, and GED is reported alongside rather than used for selection. The two are not interchangeable. Recall@$K$ measures how much of the ground truth is delivered within the top-$K$ predictions, and is insensitive to extra predicted items. GED, by contrast, charges an edit cost for every spurious node and edge. A configuration that predicts more aggressively therefore gains recall while paying in edit distance. In each of the four groups the lowest GED belongs to a setting whose recall is lower than the adopted one. Selecting on GED would trade delivered semantics for a smaller edit cost, which is the wrong trade-off for a receiver that acts on the graph it receives. The adopted setting nevertheless stays within the GED range of the main results ($0.63$ at SNR $=20$\,dB and $0.66$ channel-free, against $0.68$ and $0.66$ in Tables~\ref{tab:semantic_fidelity_snr6} and~\ref{tab:ablation_multimodal}).}
}

{\color{black}\textbf{Spatial-relation recovery.} To quantify how much navigation-relevant spatial information the O-A-R graph retains (cf.\ the scope-and-assumptions discussion in Sec.~III), we partition the relation vocabulary into spatial/directional predicates (e.g., \emph{on}, \emph{under}, \emph{above}, \emph{behind}, \emph{in front of}, \emph{next to}; $32$ of the $48$ content predicates, i.e., the $49$ predicate categories excluding the non-relational \emph{in-image} placeholder) and the remainder. As reported in Table~\ref{tab:spatial_recall}, at the high tier the model recovers spatial relations at Recall@50 \revb{$=0.58$}---comparable to its overall relation recall \revb{($0.64$)} and only modestly below the non-spatial subset \revb{($0.65$)}. This confirms that, although the graph carries no absolute coordinates, the transmitted relation layer preserves a usable \emph{relative} spatial topology, i.e., the form of spatial knowledge required for high-level embodied decisions.}

\begin{table}[!t]
\centering
\color{black}
\caption{\textbf{Spatial vs.\ non-spatial relation recovery} (high tier, SNR $=10$\,dB). Content predicates are split into spatial/directional ($32$ of $48$) and the rest.}
\label{tab:spatial_recall}
\small
\begin{tabular}{lccc}
\toprule
Relation subset & R@10 & R@20 & R@50 \\
\midrule
Spatial / directional & \revb{0.269} & \revb{0.404} & \revb{0.578} \\
Non-spatial           & \revb{0.375} & \revb{0.506} & \revb{0.649} \\
Overall (all)         & \revb{0.339} & \revb{0.478} & \revb{0.636} \\
\bottomrule
\end{tabular}
\end{table}

{\color{black}\textbf{Failure cases.} Inspecting the low-bandwidth reconstructions reveals systematic failure modes that delimit the applicable boundary of the method. (i)~\emph{Dense small objects}: when many small instances of the same class crowd a scene, semantic queries merge or omit nodes, lowering object recall. (ii)~\emph{Long-tail and open-world entities}: rare categories and out-of-vocabulary objects are frequently missed---a direct consequence of the closed-set O-A-R vocabulary. (iii)~\emph{Fine attributes at the lowest tier}: under the object-only budget, attribute and most relation edges are intentionally discarded, so attribute-dependent queries cannot be served. (iv)~\emph{Unreliable audio-derived nodes}: audio event nodes (e.g., \emph{footsteps}, \emph{chirping\_birds}, \emph{car\_horn}) are the most frequently missed or hallucinated, reflecting the weaker grounding of the audio stream relative to vision. Modes (i) and (iv) are directly illustrated in the failure columns of Fig.~\ref{fig:qualitative}\revb{(e,\,f)}: a dense classroom where \revb{three of the four chairs are missed while duplicated \emph{shelf} and \emph{leg} nodes appear} \revb{(e)}, and an ocean scene where audio-event nodes (\emph{chirping\_birds}, \emph{sea\_waves}) are hallucinated \revb{(f)}. Overall, they indicate that the framework is best suited to coarse-to-mid-grained relational perception rather than dense instance-level or open-world enumeration.}

\subsection{Downstream Task Performance: Visual Question Answering}

\begin{figure}[!t]
\centering
\includegraphics[width=3.5in]{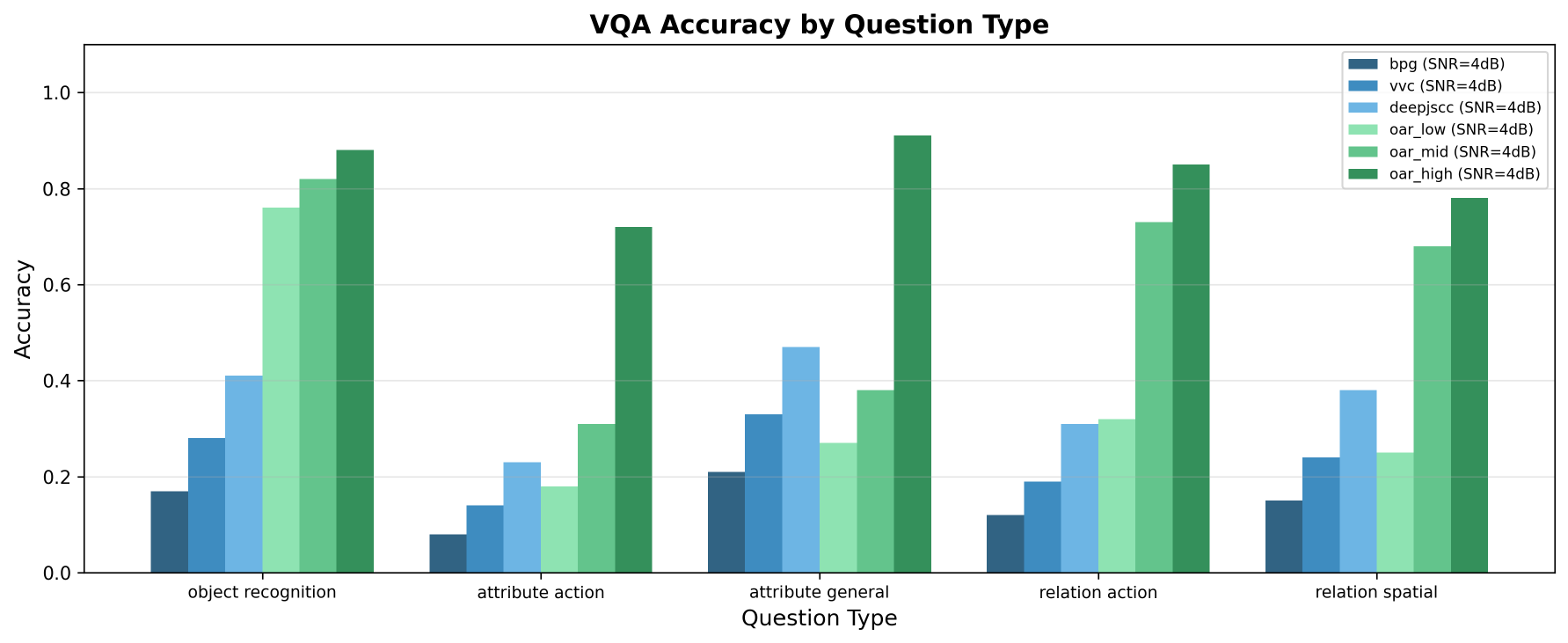}
\caption{Zero-shot VQA accuracy across different cognitive question types under deep fading channels ($\text{SNR} = 4$ dB). Our adaptive OAR framework demonstrates superior robustness and graceful degradation across Low, Mid, and High bandwidth modes compared to traditional digital codecs and DeepJSCC.}
\label{fig:vqa_results}
\end{figure}

To evaluate downstream utility and ensure fair comparisons against reconstruction-based methods, we conduct a zero-shot Visual Question Answering (VQA) task under deep fading channels ($\text{SNR} = 4$ dB). As shown in Fig.~\ref{fig:vqa_results}, traditional digital codecs (BPG, VVC) and DeepJSCC suffer severe performance collapse across all question types (accuracy $<0.4$), as pixel-level distortions cause downstream Vision-Language Models (VLMs) to hallucinate.

In contrast, our framework demonstrates robust, graceful degradation strictly aligned with embodied cognitive priorities. Under extreme constraints (Low BW), the system selectively protects fundamental entity anchors, achieving an object recognition accuracy of $\sim$0.76---nearly doubling the performance of DeepJSCC. When channel capacity improves (Mid BW), dynamic resource allocation to the relation stream triggers a significant accuracy boost in relational reasoning (e.g., spatial relation accuracy rises from $\sim$0.25 to 0.68). Finally, the High BW mode recovers the complete O-A-R graph, enabling fine-grained attribute identification (accuracy $>0.9$). This unequivocally confirms that our adaptive hierarchical strategy effectively translates extreme bandwidth savings into reliable downstream decision-making capabilities.

\section{Conclusion}

In this paper, we proposed a multimodal Object-Attribute-Relation (O-A-R) semantic communication framework \revb{for delivering decision-critical semantics to machine agents over dynamic, band-limited networks}. By abandoning pixel-level reconstruction for direct topological recovery, our system transmits decision-critical semantics via bandwidth-adaptive hierarchical scheduling and cross-modal compensation. Experiments demonstrate that at extreme low bandwidths (1--3 kbps), our approach achieves a 10-fold bandwidth saving ($>90\%$) over SOTA digital codecs and DeepJSCC. Under deep fading channels (SNR $\le$ 4 dB), the framework aligns with machine cognitive priorities, intelligently truncating fragile high-level relations to guarantee a 0\% decoding outage rate and robust object anchor perception. Furthermore, cross-modal synergy leverages audio-textual priors to compensate for severe visual degradation. Coupled with an 89\% latency reduction \revb{relative to the average baseline}, this \rev{framework offers a resilient, task-oriented solution for future machine-to-machine communications}.

\revb{\textbf{Limitations and future work.} We close with a critical assessment of the constraints that bound the applicability of the present framework. (i)~\emph{Closed-set, domain-bound vocabulary.} The O-A-R vocabulary is fixed at training time and derived from Visual Genome; open-world and long-tail entities are missed, and deployment in a new domain requires re-training or vocabulary adaptation. (ii)~\emph{Semantic-level tasks only.} Because no pixel evidence is delivered, tasks that require photometric or geometric detail---surface-defect inspection, texture assessment, or human viewing---cannot be served; the framework is a complement to, not a replacement for, general-purpose video coding. (iii)~\emph{Idealized channel knowledge.} The adaptive controller assumes a reliable CSI feedback link and receiver-side perfect-CSI equalization; feedback delay, quantization, and estimation error are not yet modeled, and training covers a matched $[0,20]$\,dB regime, so generalization to unseen channel statistics is not guaranteed. (iv)~\emph{Single-frame operation.} The system transmits per-snapshot semantics and does not yet exploit temporal redundancy across video frames. (v)~\emph{No absolute localization.} The graph carries only relative spatial predicates; metric localization would require an additional coordinate-regression head. (vi)~\emph{Residual failure modes.} As analyzed in Sec.~IV, dense same-class instances and weakly grounded audio-derived nodes remain systematic error sources. Addressing these boundaries---in particular temporal extension, open-vocabulary O-A-R modeling, and robust operation under imperfect CSI---constitutes our future work.}

\bibliographystyle{IEEEtran}

\bibliography{ref}

\begin{IEEEbiography}[{\includegraphics[width=1in,height=1.25in,clip,keepaspectratio]{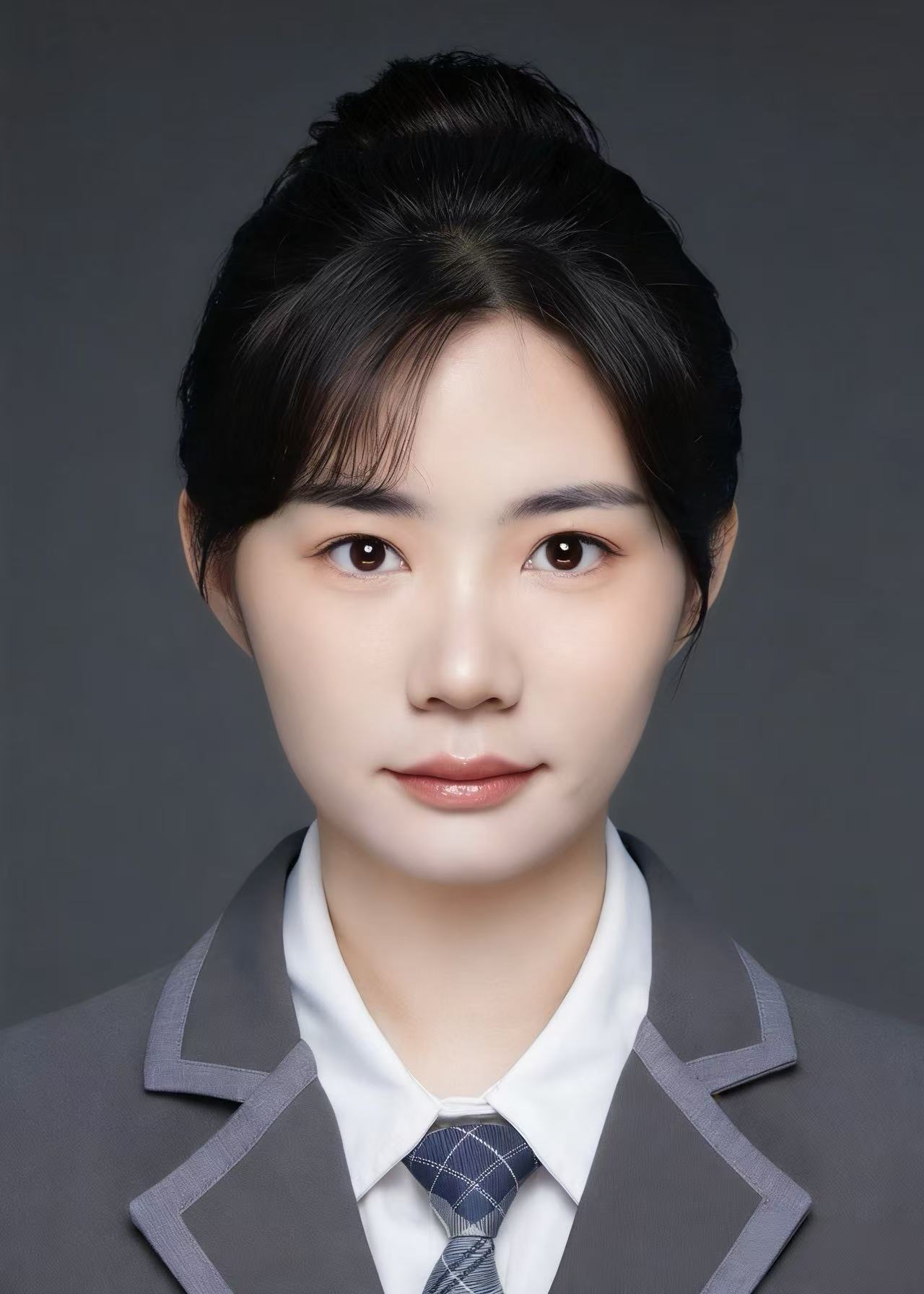}}]{Chenxing Li}
(Student Member, IEEE) received a B.E. degree from Beijing University of Posts and Telecommunications, Beijing, China, in 2018, and received a master degree in Electronic Engineering from Tsinghua University, Beijing, China, in 2024. She is currently working towards her Ph.D. degree at Tsinghua University. Her current research interests include machine learning, computer vision, and multimedia communications.
\end{IEEEbiography}

\begin{IEEEbiography}[{\includegraphics[width=1in,height=1.25in,clip,keepaspectratio]{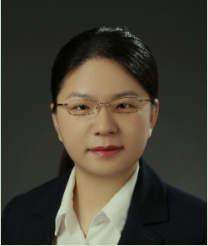}}]{Yiping Duan}
(Member, IEEE) received the Ph. D. degree from the school of computer science and technology, Xidian University in 2016. She is currently an associate research fellow with the Department of Electronic Engineering, Tsinghua University. Her current research interests include multimedia communications, video codecs, and multimedia signal processing.
\end{IEEEbiography}

\begin{IEEEbiography}[{\includegraphics[width=1in,height=1.25in,clip,keepaspectratio]{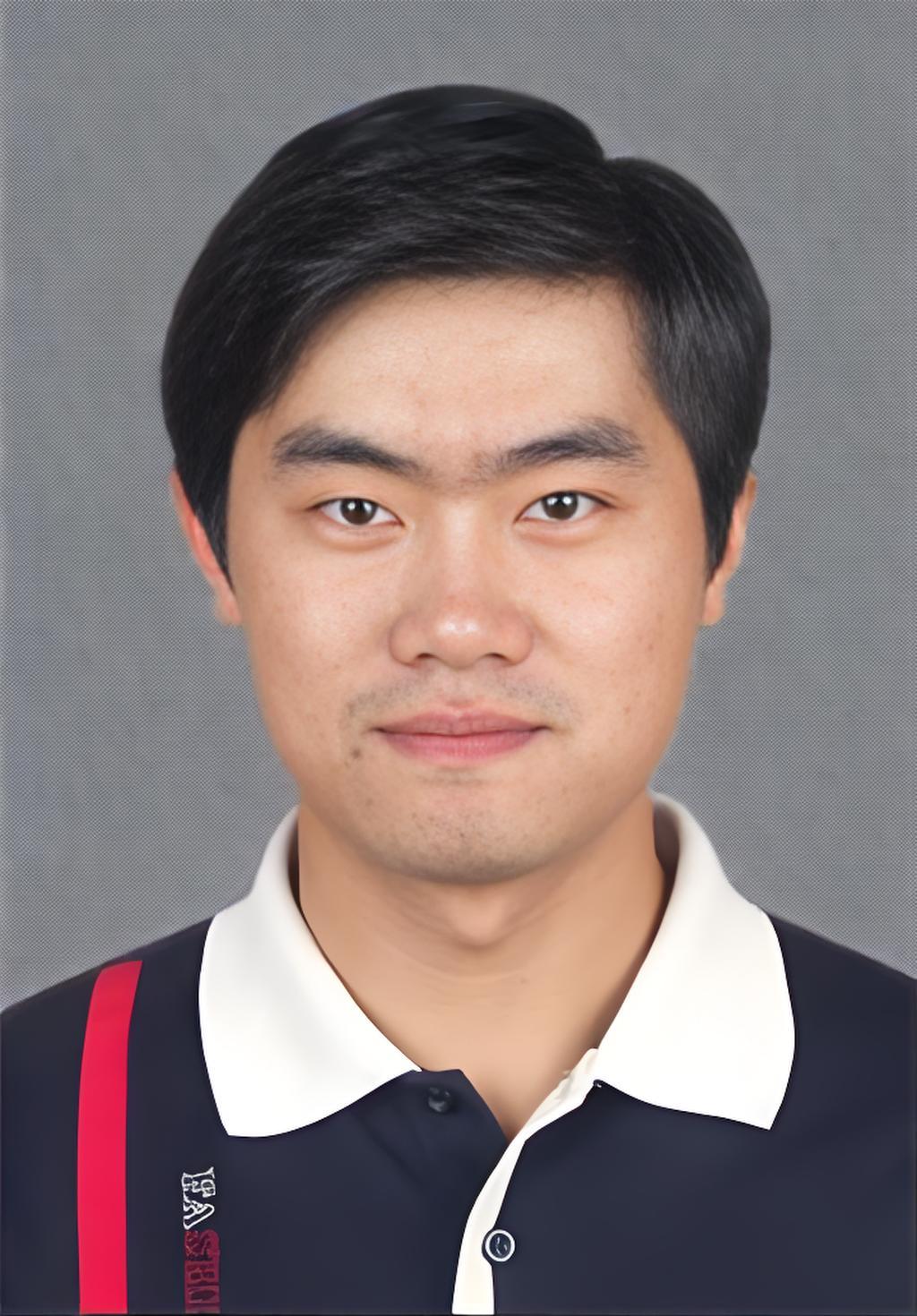}}]{Han Jiao}
(Member, IEEE) received the Ph.D. degree in computer and information science from the University of South Australia. He is currently an associate research fellow with the Department of Electronic Engineering, Tsinghua University, and also affiliated with the National Key Laboratory of Space Network and Communication. His expertise spans complex system design, system integration, big data analytics, and their applications in border and coastal defense, smart cities, and low-altitude security. His current research focuses on situational awareness and intelligent computing in complex environments.
\end{IEEEbiography}

\begin{IEEEbiography}[{\includegraphics[width=1in,height=1.25in,clip,keepaspectratio]{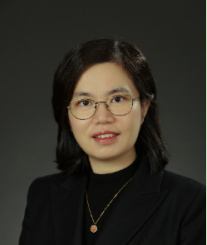}}]{Xiaoming Tao}
(Senior Member, IEEE) received a Ph.D. degree in information and communication systems from Tsinghua University, Beijing, China, in 2008. She is currently a Professor with the Department of Electronic Engineering at Tsinghua University. Prof. Tao served as a workshop General Co-Chair for IEEE INFOCOM 2015 and the volunteer leadership for IEEE ICIP 2017. She has been the Editor of the Journal of Communications and Information Networks and China Communication since 2016. She is also the recipient of the National Science Foundation for Outstanding Youth (2017--2019) and many national awards, e.g., the 2017 China Young Women Scientists Award, 2017 Top Ten Outstanding Scientists and Technologists from China Institute of Electronics, 2017 First Prize of Wu Wen Jun AI Science and Technology Award, 2016 National Award for Technological Invention Progress, and 2015 Science and Technology Award of China Institute of Communications, etc.
\end{IEEEbiography}

\begin{IEEEbiography}[{\includegraphics[width=1in,height=1.25in,clip,keepaspectratio]{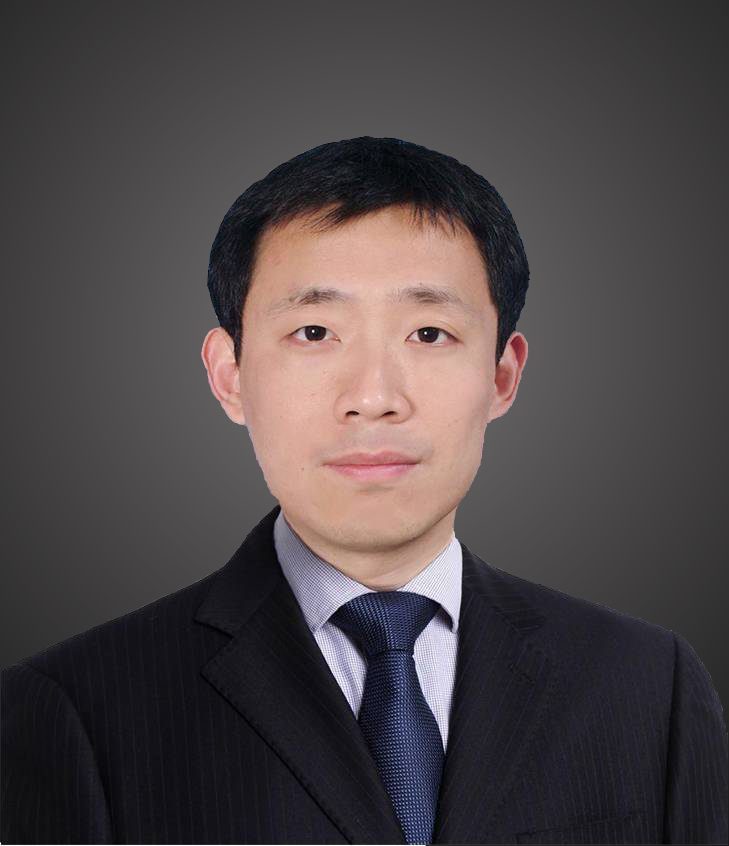}}]{Weiyao Lin}
(Senior Member, IEEE) received the BE and ME degrees from Shanghai Jiao Tong University, Shanghai, China, in 2003 and 2005, respectively, and the PhD degree from the University of Washington, Seattle, WA, USA, in 2010, all in electrical engineering. He is currently a professor with the Department of Electronic Engineering, Shanghai Jiao Tong University. He has authored or coauthored more than 100 technical articles on top journals/conferences including the IEEE Transactions on Pattern Analysis and Machine Intelligence, the International Journal of Computer Vision, the IEEE Transactions on Image Processing, CVPR, NeurIPS, ICLR, and ICCV. He holds more than 20 patents. His research interests include video/image analysis, computer vision, and video/image processing applications.
\end{IEEEbiography}

\begin{IEEEbiography}[{\includegraphics[width=1in,height=1.25in,clip,keepaspectratio]{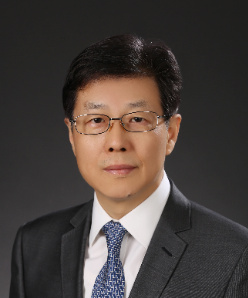}}]{Mingquan Lu}
(Senior Member, IEEE) received the M.E. and Ph.D. degrees in electrical engineering from the University of Electronic Science and Technology of China, Chengdu, China, in 1993 and 1999, respectively. He is a Professor with the Department of Electronic Engineering, Tsinghua University, Beijing, China. He also directs the Positioning, Navigation and Timing (PNT) Research Center, which develops GNSS and other PNT technologies. His current research interests include GNSS system modeling and simulation, signal design and processing, and receiver development. Dr. Lu is an ION Fellow. He was a recipient of the 2020 ION Thurlow Award.
\end{IEEEbiography}

\end{document}